\documentclass[aps,prb,twocolumn,superscriptaddress]{revtex4-2}

\usepackage{amsmath, amssymb,  graphicx, caption, subcaption}

\usepackage[colorlinks=true ,urlcolor=blue,urlbordercolor={0 1 1}]{hyperref}

\usepackage{color}
\usepackage{xcolor}
\usepackage{braket}

\usepackage[utf8]{inputenc}
\usepackage{bm}
\usepackage{verbatim}
\usepackage{graphicx}
\usepackage{amsmath}
\usepackage{color}
\usepackage{float}
\usepackage{bbm}
\usepackage{amssymb}
\usepackage{slashed}
\usepackage{wasysym,bm,bbm,dsfont}

\begin{document}

\title{
Exciton and light induced ferromagnetism  from doping a  moir\'e Mott insulator}
\author{Hui Yang}
\author{Ya-Hui Zhang}
\affiliation{Department of Physics and Astronomy, Johns Hopkins University, Baltimore, Maryland 21218, USA}
\date{\today}
\begin{abstract}

Significant efforts have been dedicated to achieving excitonic insulators. In this paper, we explore a new problem of doping excitons into a Mott insulator instead of a band insulator. Specifically, we start with a Mott insulator on a triangular moiré superlattice in a transition metal dichalcogenides (TMD) layer and inject excitons by either transferring particles to a different layer or optically pumping electrons from the valence to the conduction band. In both cases, the excitons move in the presence of local spin moments inherited from the Mott insulator. When the Heisenberg spin coupling $J$ is small, the kinetic energy of the excitons decides the magnetism, akin to Nagaoka ferromagnetism in hole-doped Mott insulators. Through density matrix renormalization group (DMRG) calculations, we demonstrate that the spin moments originating from the Mott insulator form $120^\circ$ antiferromagnetic or ferromagnetic order for the two signs of the exciton hoppings over a broad range of exciton densities. Notably, the optical pump case may result in an antiferromagnetic to ferromagnetic transition with increasing exciton density, indicating a potential mechanism for light-induced ferromagnetism. A similar exciton-induced ferromagnetism could be achieved in a moiré-monolayer system where the monolayer is electron-doped while the moiré Mott insulator is hole-doped. Our works demonstrates a new possibility to engineering magnetism through doping neutral excitons.
\end{abstract}

\maketitle

\textbf{Introduction} Excitonic insulators have been the focus of much research attention over the past few decades\cite{jerome1967excitonic,zittartz1967theory,halperin1968excitonic,comte1982exciton}. Recent experimental advances in two-dimensional materials have opened up new avenues for investigating exciton physics\cite{eisenstein2014exciton,li2017excitonic,liu2017quantum,wang2018colloquium,kogar2017signatures,wang2019evidence,wilson2021excitons,jia2022evidence,regan2022emerging}. In this article, we will explore a new direction where excitons are doped into a Mott insulator instead of a simple band insulator. Mott insulators have been known to host fascinating quantum phenomena such as frustrated magnetism and quantum spin liquids\cite{balents2010spin}, owing to the presence of localized spin moments. Doping a Mott insulator with fermionic charge carriers has been a major focus of modern condensed matter physics due to its connection to the  high-temperature superconducting cuprates\cite{lee2006doping}. However, doping a  Mott insulators with bosonic charge carriers has been comparatively underexplored, both theoretically and experimentally. Despite this, the interplay between the mobile carriers and the localized spin moments still suggests the possibility of intriguing quantum phases for bosonic carriers.  In recent years, significant advancements in experimental techniques have enabled the injection of excitons into a moir\'e Mott insulator based on transition metal dichalcogenide (TMD) bilayers\cite{gu2022dipolar,zhang2022correlated,xu2022tunable,wang2022light,chen2022excitonic,zhang20214,zhang2022,huang2022mott} As a result, it has become increasingly crucial to develop theoretical models and predictions for exciton-doped Mott insulators, which is precisely what we aim to accomplish in this paper.

Moir\'e superlattices are exceptional platforms for investigating strongly correlated physics\cite{cao2018correlated,cao2018unconventional,andrei2021marvels,mak2022semiconductor}. Among these, moir\'e superlattices based on TMD hetero-bilayer and homo-bilayer have shown promising results in simulating Hubbard model physics\cite{wu2018hubbard,Wu2019,zhang2020moire,pan2020quantum,pan2021interaction,zang2022dynamical}. A plethora of experimental discoveries have already been made, including Mott insulator\cite{Tang2020,Regan2020}, generalized Wigner crystal\cite{Regan2020,li2021imaging}, continuous metal insulator transition\cite{li2021continuous,Ghiotto2021}, quantum anomalous Hall effect\cite{Li2021}, and Kondo physics\cite{zhao2023gate}. Building on the existing research, we now investigate the effects of doping neutral excitons into the Mott insulator.  This can be achieved through two methods: transferring particles to a different layer using a displacement field in a bilayer system\cite{gu2022dipolar,zhang2022correlated}, or optically pumping electrons from the valence band to the conduction band\cite{regan2022emerging,wang2022light,Jin2019,xiong2022bosonic,Seyler2019,Tran2019,Alexeev2019,Zhang2021,gao2023excitonic}. We will primarily focus on the former approach, as the equilibrium excitons in this case do not have a limited lifetime. However, it is important to note that our model and predictions also apply to optically pumped excitons within their lifetime. Many previous studies have used excitons to probe correlated states\cite{Miao2021,PhysRevLett.127.037402,Shimazaki2020,Zhou2021}, but in this work, we focus on the novel physics that arises from a finite density of excitons.

Let us consider a bilayer system consisting of a moir\'e layer at the bottom and a monolayer TMD at the top, separated by an insulating hBN barrier. Initially, the system has $n_b=1$ and $n_t=0$, resulting in the moir\'e layer being in a Mott insulating phase. Our first step is to obtain Wannier orbitals of the exciton by solving Schrodinger's equations with one single vacancy or two nearby vacancies in the moir\'e layer. This approach leads to an effective low-energy spin-exciton model with four states on each moir\'e site: (1) one hole in the moir\'e layer with spin up\footnote{The particle can be electron or hole depending whether we dope into the conduction or valence band.}; (2) one hole in the moir\'e layer with spin down; (3) one exciton with spin up in the monolayer; and (4) one exciton with spin down in the monolayer. It is important to note that a hole and an exciton can not simultaneously occupy the same site. Our model involves four crucial parameters: the hopping of the excitons ($J_p$), the repulsive interaction between two nearby excitons ($J_{pz}$), the spin-spin coupling in the monolayer ($J_t$), and the spin-spin coupling in the moir\'e layer ($J_b$). Specifically, we are interested in the regime with $n_b=1-x$ and $n_t=x$, where $x$ denotes the exciton density. Our model is particularly suited for the small $x$ regime, where the excitons remain stable against dissociation into an electron hole gas.

We then employ density matrix renormalization group\cite{tenpy} (DMRG) technique to simulate the model. Our focus is on the regime where $J_b$ is small or zero, which is a reasonable assumption for a strong Mott insulator with a large $U/t$, as is demonstrated in the TMD hetero-bilayer\cite{Tang2020}. In this regime, the magnetism of the localized spin moments in the moir\'e Mott layer is primarily governed by the kinetic energy of the excitons, akin to the Nagaoka ferromagnetism\cite{nagaoka1966ferromagnetism} and kinetic antiferromagnetism\cite{haerter2005kinetic,davydova2022itinerant,morera2022high,lee2022triangular} in the hole doped Mott insulator, which is different from the magnetic order controlled by spin-orbit coupling studied in Ref.~\onlinecite{PhysRevResearch.4.043126}. Specifically, we observe a 120$^\circ$ antiferromagnetic order or a ferromagnetic order for $J_p>0$ and $J_p<0$, respectively, where the sign of exciton hopping ($J_p$) plays a pivotal role. The later is realized if we hole-dope the moir\'e layer with $n^h_b=1+x$ holes in the valence band while electron-doping the monolayer with $n^e_t=x$ electrons in the conduction band.  In this case there is an antiferromagnetic to ferromagnetic transition when increasing the exciton density $x$ if there is a finite but small $J_b$ at $x=0$. The same conclusion holds true in the case of optically pumped excitons, offering a plausible mechanism for light-induced ferromagnetism. This picture may also be relevant to the experimental findings in Ref.~\onlinecite{wang2022light}.

\begin{figure}[ht]
\centering
\includegraphics[width=0.4\textwidth]{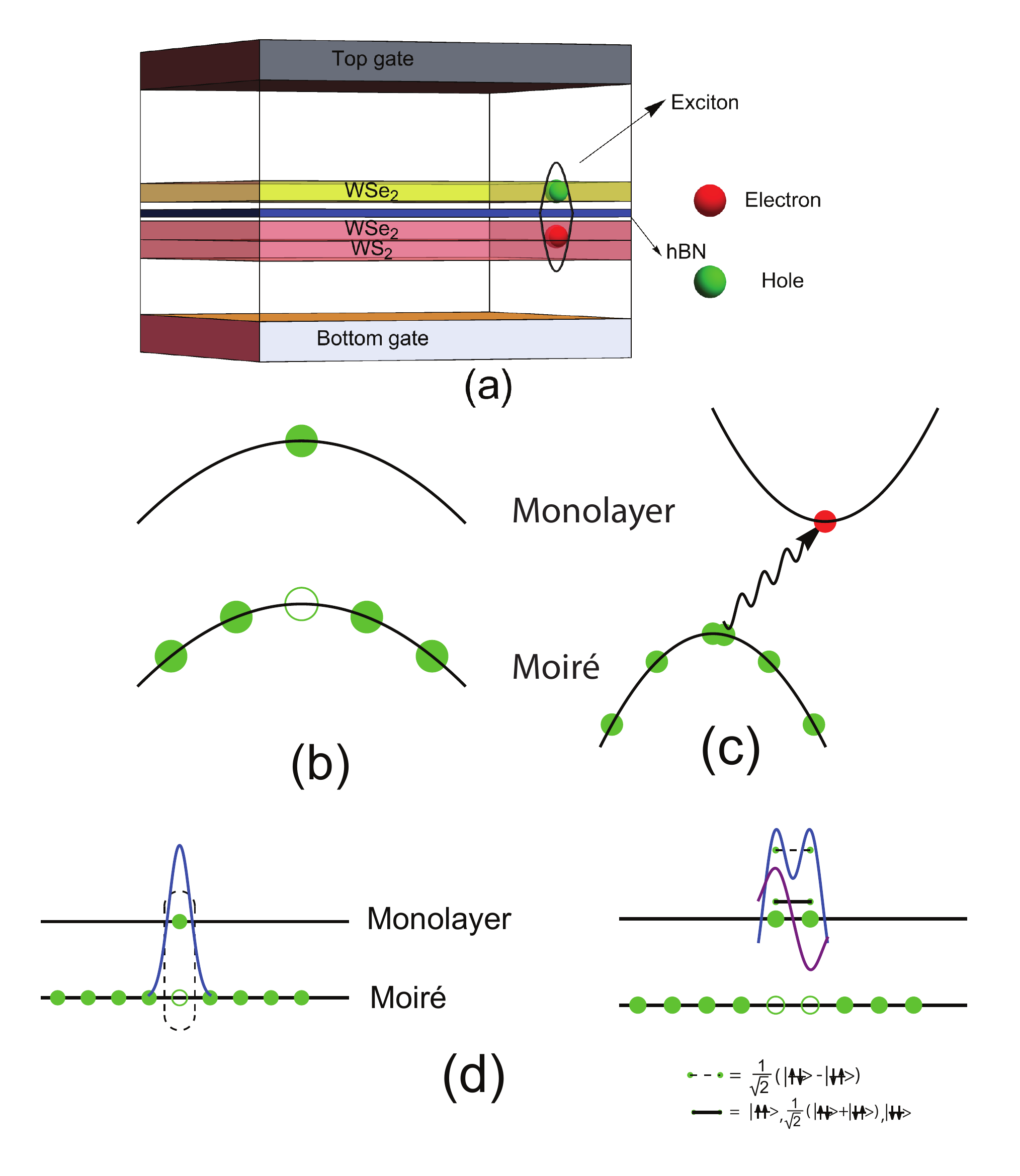}
\caption{ (a)  An illustration of the moir\'e+monolayer system.  (b)(c) Two different ways of doping excitons into a Mott insulator. Here, solid red circle denotes an electron in the conduction band, and empty red labels a vacancy of electron in the conduction band. Similarly solid Green circle means a hole in the valence band, while empty Green circle means a vacancy of hole in  the valence band. A vacancy of hole is equivalent to an electron in the valence band, but we will call it vacancy of hole to be distinguished from an electron in the conduction band. (b) $n^h_b=1-x, n^h_t=x$. (c)$n^h_b=1+x, n^e_t=x$.
(d) Illustration of the wave function of the holes in the top layer when there is one vacancy (on the left), and two vacancies (on the right) doped into the moir\'e Mott insulator in the bottom.  When there are two vacancies nearby, the two holes form a spin singlet or a spin triplet, whose energy difference gives the parameter $J_t$ in our model in Eq.~\ref{eq2}.  }
\label{fig1}
\end{figure}

\textbf{Model} We consider the experimental setup shown in Fig.~\ref{fig1}(a). We have a moir\'e superlattice formed by WSe$_2$/WS$_2$ hetero-bilayer in the bottom, separated from a WSe$_2$ monolayer in the top by an insulating hexagon boron nitride (hBN) barrier.   We call this setup moir\'e+monolayer. Because inter-layer tunneling is suppressed by the hBN barrier, the top and bottom WSe$_2$ layer have separately conserved total number of charges.   Both the  moir\'e layer in the bottom  and the monolayer in the top  can be either electron or hole doped from the charge neutrality, which is a band insulator with large band gap at order of eV. To avoid confusion, we define electron (or hole) density per moir\'e unit cell of the layer $a=t,b$ as $n^e_a$ (or $n^h_a)$.  The total charge density at layer $a$ is thus $\rho_a=n^h_a-n^e_a$. Our starting point is $n^h_b=1, n_t=0$, where the bottom moir\'e layer forms a Mott insulator with one hole per moir\'e site while the top layer is still at charge neutrality.  Then we dope excitons in two different ways:  (I) We tune the density to be $n^h_b=1-x, n^h_t=x$, shown in Fig.1(b); (II) We tune the density to be $n^h_b=1+x, n^e_t=x$, shown in Fig.1 (c).   In the first case, we have excitons formed by a vacancy of hole in the valence band of the bottom layer and a hole in the valence band of the top layer. In the second case, we have excitons formed by an additional hole in the valence band of the bottom layer and an electron in the conduction band of the top layer.  We will see later that these two cases can be captured by similar effective model, but with opposite signs of the hopping of excitons, which will qualitatively change the magnetic physics of the localized spin moments in the bottom moir\'e layer.  Both these cases can be achieved in equilibrium with a displacement field $D$ in a dual gated sample to tune the densities of the top and bottom layers separately.  The second case can also be realized in non-equilibrium through optical pumping.

We focus on the case I as an example to derive a low energy effective model to capture the exciton and magnetic moments.  We start from the Mott insulator in the moir\'e layer, which is captured by an extended lattice Hubbard model \cite{wu2018hubbard,Wu2019,zhang2020moire,pan2020quantum,pan2021interaction,zang2022dynamical}:
\begin{align}
    H_b=-t_b\sum_{<ij>}h^\dag_{bi}h_{bj}+\frac{U}{2}\sum_{i}n_{bi}^2+V\sum_{<ij>}n_{bi}n_{bj}.
    \label{eq1}
\end{align}
From our calculation (see the supplementary), we get $t_b=1.06$meV, $U=660.635/\epsilon$ meV and $V=92.689/\epsilon$ meV, where $\epsilon$ is the renormalization factor of the dielectric constant which should be at order of  $10$. 

\begin{figure}[htbp]
\centering
\includegraphics[width=0.4\textwidth]{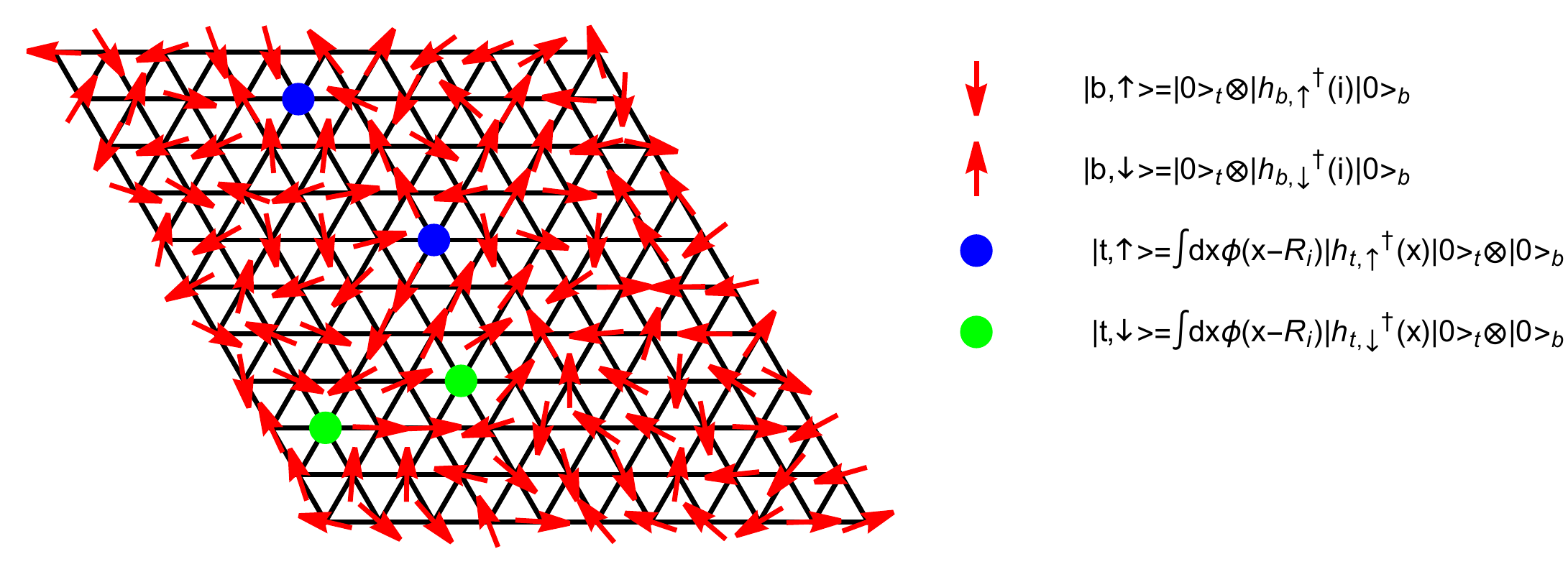}
\caption{Illustration of the moir\'e lattice and the four-dimensional local Hilbert space. The local moments represent the spin degree of freedom in the bottom layer. The solid disks represent the doped excitons with density $x$. Different colors correspond to different spins in the top layer of  the exciton state. If we polarize the spin in the top layer, there is one flavor for exciton and the model reduces to a bosonic version of  $t-J$ model.}
\label{fig2}
\end{figure}

The Mott insulator  has one hole per moir\'e site in the bottom layer.  When the system is doped with one hole in the top layer and one vacancy of hole (electron) in the bottom layer, the hole in the top layer will be trapped to the vacancy of hole (which is an electron) in the bottom layer, leading to a neutral exciton.  Let us treat the $t_b$ term as a small perturbation and ignore it for now. Then the vacancy in the bottom layer is fixed at one site $i$. The hole in the top layer moves under the potential $\sum_{j\neq i}V_{j}(x)$, where $V_j(x)$ is the Coulomb interaction between the hole in the top layer and the hole at moir\'e site $j$ of the bottom layer.  We can get the  wave function $\psi_{i}(x-{\bf R}_i)$ for the hole in the top layer by solving the corresponding Shr\"odinger equation. As schematically shown in Fig.~\ref{fig1}(d), the hole wave function is very localized around the vacancy site ${\bf R}_i$. Combining the two spins of the hole in the top layer,  we have two exciton states $\ket{t,\sigma}=\int dx \psi_i(x-R_i) h_{t\sigma}^\dagger(x) \ket{0}_t \ket{0}_b$ for each site $i$. Here $\sigma=\uparrow,\downarrow$.  We have two additional states without exciton at site $i$: $\ket{b,\sigma}=h^\dagger_{i;b \sigma}\ket{0}_b \otimes \ket{0}_t$. Again $\sigma=\uparrow,\downarrow$.   These four states are shown in Fig.~\ref{fig2}.  Because of the strong inter-layer repulsion, doubly occupied states with holes in both top and bottom layer at the same site is penalized and ignored.

\begin{figure}[htbp]
\centering
\includegraphics[width=0.4\textwidth]{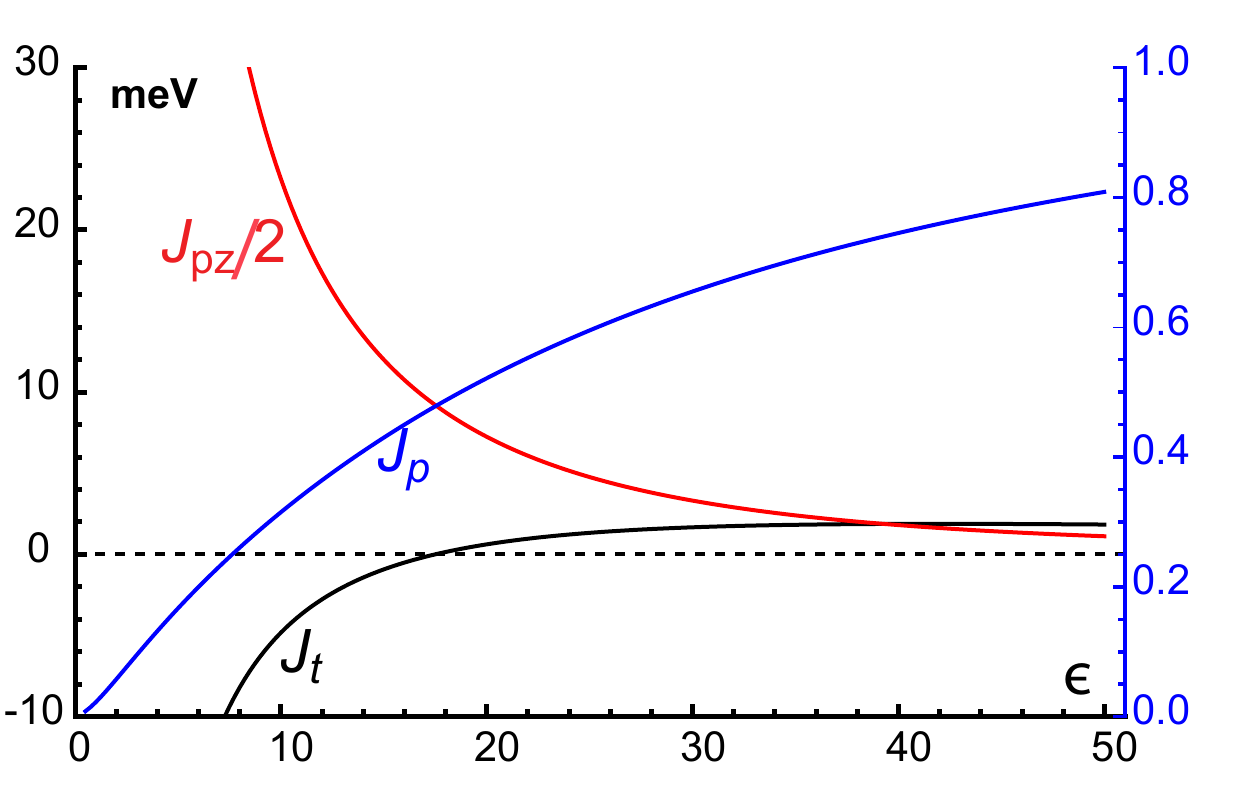}
\caption{The couplings in our effective model as a function of the dielectric constant $\epsilon$. The vertical axis on the left is for $J_{pz}$ and $J_t$, and the vertical axis on the right is for $J_p$. In our calculation, the lattice constants are $0.328nm$ for WSe$_2$, $0.315nm$ for WS$_2$, and the moir\'e lattice constant is $a_M=7.9nm$ at zero twist angle. $m=0.42m_e$ is the effective mass for the valence band of WSe$_2$.  The distance of the top layer and bottom layer is $d=2nm$.   }
\label{fig3}
\end{figure}

Now we can proceed to derive the effective hopping of the exciton. The four states at each site can be constructed from a tensor product of a layer pseudospin $1/2$ $\vec P$ and the real spin $\vec S $ \cite{zhang2022}. Then the exciton creation and annihilation operator correspond to $P^\dagger$ and $P^{-}$. The spin operator in the top layer is $\vec S_t(i)=\frac{1}{2}(1+P_z)\vec S(i)$, while the spin operator in the bottom layer is $\vec S_b(i)=\frac{1}{2}(1-P_z)\vec S(i)$. $\vec P_z(i)$ is the layer polarization or equivalently a dipole moment.  $1+P_z(i)$ is the exciton occupation number at the site $i$.  With these operators, an effective four-flavor model can be written down:

\begin{align}   H=&\sum_{<ij>}J_t\vec{S}_t(i)\cdot\vec{S}_t(j)+J_b\vec{S}_b(i)\cdot\vec{S}_b(j)+\frac{1}{2}J_{pz}P_z(i)P_z(j)\nonumber\\
    +&\frac{1}{2}J_p(P_x(i)P_x(j)+P_y(i)P_y(j))(4\vec{S}(i)\cdot\Vec{S}(j)+S_0(i)S_0(j)),
    \label{eq2}
\end{align}
where $J_p$ is the exciton hopping. The exciton hopping corresponds to hole in the top layer and electron in the bottom layer hopping simultaneously. Its hopping parameter can be obtained from first order perturbation of the $t_b$ term and we get $J_p=t_b\int\psi_i^*(x-R_i)\psi_j(x-R_j)$ as the exciton hopping.   $J_{pz}/8$ gives the dipole-dipole repulsion for two nearby excitons. $J_b$ is the super-exchange spin coupling of the bottom layer, which we assume is small in the strong $U_b/t_b$ limit. $J_t$ is the spin-spin coupling in the top layer, which has two competing contributions: Hund's coupling to favor spin-triplet and covalent bonding to favor spin-singlet. The evolution of these parameters with the dielectric constant $\epsilon$ is shown in Fig.~\ref{fig3}. When $\epsilon=20$, we have $J_{pz}/2=7.25075meV$, $J_{p}=0.51646meV$, $J_t=0.621032meV$, here $J_{p}=0.487 t_b$  here we use $t_b=1.06meV$ as derived from Wannier orbital construction. $J_b=\frac{4 t_b^2}{U}\approx 0.14$ meV and is smaller than other values.

So far we discussed the case I. In the case II, we dope an additional hole in the bottom layer, accompanied by an electron in the conduction band of the top layer. The physics is described by the same model as in the case I, except now the hopping term $J_p$ is negative. The effective hopping of an exciton is from a second order process of hopping electron and hole. In the case I electron and hole are from the same valence band, while in the case II they are from conduction and valence band. This gives a sign difference.

\textbf{Reduction to a bosonic t-J model} Let us assume that the spin of the top layer is polarized to the spin up. Then the four states at each site reduces to three states, we can label them $\ket{\uparrow}=\ket{b,\uparrow},\ket{\downarrow}=\ket{b,\downarrow}, \ket{0}=\ket{t,\uparrow}$. One can see that the Hilbert space is similar to the familiar $t$-$J$ model with the exciton state playing the role of the empty site.  The difference is that the doped carriers are neutral and bosonic. Nevertheless, we expect the influence to the magnetism may be similar to the fermionic t-J model, especially when $x$ is small. We will confirm this picture below.

\textbf{Exciton induced magnetism} When the exciton density $x$ is zero, the magnetic physics in the bottom layer is governed by the super-exchange Heisenberg coupling $J_b\approx \frac{4 t_b^2}{U}$. We will focus on the regime where $U>>t_b$ and $J_b$ is very small, as observed in the TMD hetero-bilayer at zero twist angle. Then at $x=0$ all of the spin configurations are degenerate.  Here we are interested in how the movement of excitons influence the magnetic ordering at finite $x$. 

We will see that the spin $\vec S_t$ in the top layer is spin polarized. Then as said before the physics is captured by a bosonic version of $t-J$ model. If there is only one single exciton, then the statistics does not matter and it is equivalent to the familiar fermionic $t-J$ model with a single hole. So we can quote the previous studies of the fermionic t-J model to understand our system.  In the single hole doped Mott insulator, it is known that the magnetic ordering in the infinite U limit is decided by the kinetic energy of the holes, which leads to either the 120$^\circ$ antiferromagnetic order\cite{haerter2005kinetic,davydova2022itinerant,morera2022high,lee2022triangular} or the Nagaoka ferromagnetic order (FM) depending on the sign of the hopping\cite{nagaoka1966ferromagnetism}.   Following this mapping, we reach the conclusion that the magnetic order of $\vec S_b$ is $120^\circ$ AFM if $J_p>0$ and spin polarized if $J_p<0$ for the single exciton case.

To check whether the conclusion holds for finite densities of excitons,  we adopt infinite density matrix renormalization group (iDMRG) to simulate the model Eq.~\ref{eq2}. The iDMRG simulation is performed on $L_x\times L_y=6\times6$ cylinder. $L_x$ is along the direction ${\bf a}_1=(1,0)$ and $L_y$ is along the direction ${\bf a}_2=(-\frac{1}{2},\frac{\sqrt{3}}{2})$. The bond dimension is up to $5000$ and the truncation error is $10^{-5}$.  The exciton density $x$ is from $\frac{1}{18}$ to $\frac{17}{18}$.\\

\begin{figure}[h]
\centering
\includegraphics[width=0.5\textwidth]{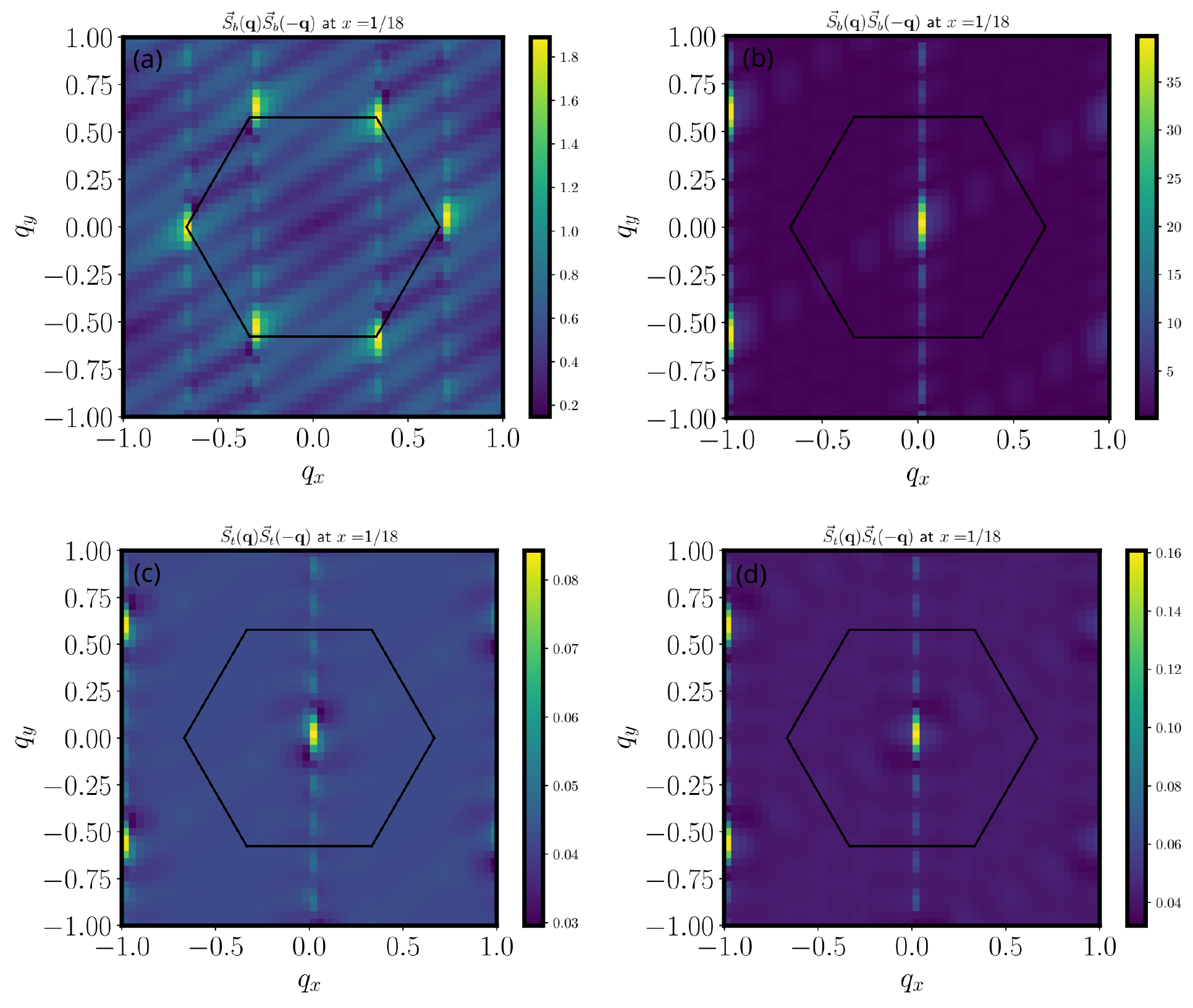}
\caption{The spin-spin correlation function for fixed $J_b=J_t=0$, $J_{pz}=5$, $x=\frac{1}{18}$. (a) and (b) are the spin correlation function $ \langle\vec{S}_b({\bf q})\vec{S}_b(-{\bf q})\rangle$ in the bottom layer, and (c), (d) are the spin correlation function $\langle\vec{S}_t({\bf q})\vec{S}_t(-{\bf q})\rangle$ in the top layer. (a) and (c) results for $J_p=1$. (b) and (d) results for $J_p=-1$.}
\label{fig4}
\end{figure}

We first fix $J_b=J_t=0$ and $J_{pz}=5$, so the spin configurations are selected purely by the exciton hopping term $J_p$. The spin structure factors are shown in Fig.~\ref{fig4}. At $x=\frac{1}{18}$, for $J_p=1$, in Fig.~\ref{fig4}(a) and (c). We can read that the spin in the bottom moir\'e layer is $120^\circ$ ordered, while the spin in the top layer is ferromagnetically ordered. In contrast, if $J_p=-1$, as shown in Fig.~\ref{fig4}(b) and (d), the spin in both layers are now FM ordered.  The spin configuration in the bottom moir\'e layer exactly follows our expectation from analog to the hole doped case.  There is an intuitive explanation in Schwinger boson mean field theory which we list in the supplementary. Once the magnetic order in the bottom layer is fixed, the exciton only carries the spin $1/2$ in the top layer and we have spinful boson gas with density $x$, which is known to be in a spin polarized Bose-Einstein condensation (BEC) phase. This explains the FM order of the top layer.

\begin{figure*}[htbp]
\centering
\includegraphics[width=1\textwidth]{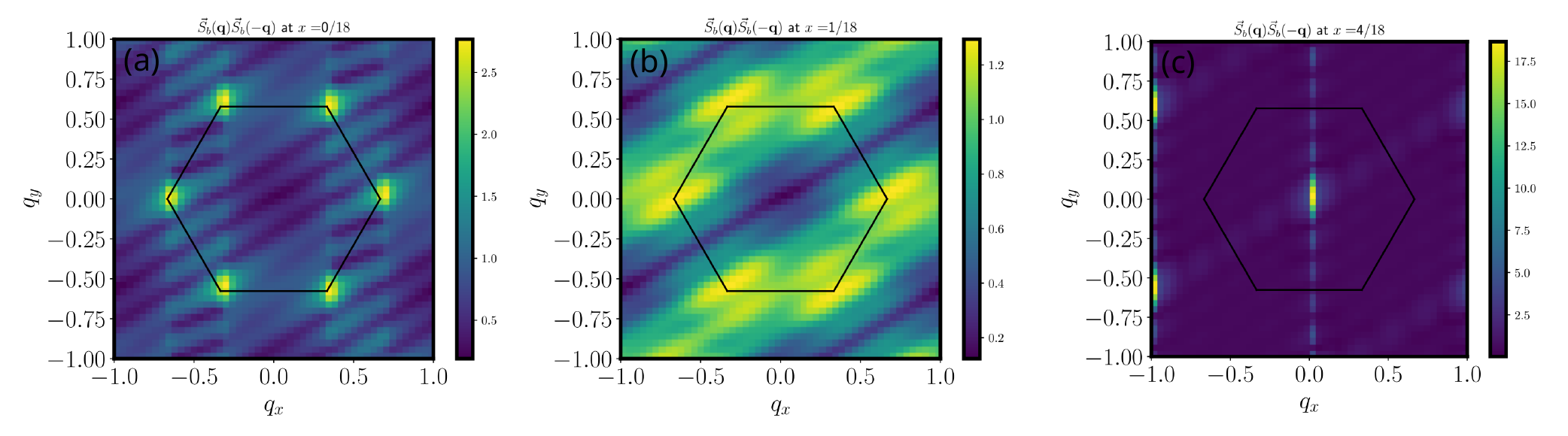}
\caption{(a) and (b) the spin correlation function $\langle\vec{S}_b({\bf q})\vec{S}_b(-{\bf q})\rangle$ for fixed $J_t=0$, $J_{pz}=5$, $J_p=-1$, $J_b=0.06$. (a), (b), (c) correspond to the increasing of exciton density.}
\label{fig5}
\end{figure*}

The case of $J_p<0$ is particularly interesting as the FM order is quite robust to large exciton density (see the supplementary).  $J_p<0$ can be realized in the case II, with electron and hole doped into the conduction and valence band of the two layers respectively. In real systems there may be a  small but finite $J_b>0$. Then there is a competition between the AF order from $J_b$ and the kinetic FM.  In Fig.~\ref{fig5}, we show a transition from an antiferromagnetic state to a ferromagnetic state in the bottom layer as we increase the exciton density, while the spin in the top layer is always polarized during this transition (see the supplementary). In the FM phase we also have exciton condensed at momentum $Q=0$, forming a spin polarized superfluid (see the supplementary).  Note that the case II can also be realized in optical pumping, with exciton density proportional to optical power. Hence this provides a mechanism of light induced ferromagnetism.  Light induced FM  was  observation recently\cite{wang2022light}  at fractional total filling $n=-\frac{1}{3}$. Similar model can be shown to describe exciton doped Wigner crystal at $n=-\frac{1}{3}$ (see the supplementary) and thus our theory may offer a natural explanation of the observation in Ref.~\onlinecite{wang2022light} in terms of kinetic driven ferromagnetism.

\textbf{Conclusion} In summary, we study exciton doped Mott insulator in the TMD moir\'e systems. In these systems the spin-spin coupling $J$ of the Mott insulator is usually very small because $U/t$ is very large. As a result, the magnetism of the localized moments will be decided by the kinetic term of the exciton. We perform a DMRG simulation and find that the spin moments inherited from the Mott insulator form $120^\circ$ AF order or ferromagnetic order depending on the sign of the effective hopping of excitons. Especially, if we dope a moir\'e+monolayer with electron and hole in the two layers respectively, there can be an antiferromagnetic to ferromagnetic transition when increasing the exciton density. The same physics can happen through optical pumping, providing a mechanism of light induced ferromagnetism. Our work demonstrates the possibility of engineering magnetism through doping neutral excitons. 

\textbf{Acknolwedgement} YHZ thanks Mohammad Hafezi, Tsung-Sheng Huang, Yi Li and Feng Wang for discussion. This work was supported by the National Science Foundation under Grant
No. DMR-2237031. The iDMRG simulation was performed using the TeNPy
Library(version 0.10.0)\cite{tenpy}. The numerical simulation was carried out at the
Advanced Research Computing at Hopkins (ARCH) core facility (rockfish.jhu.edu), which is supported by the National
Science Foundation (NSF) grant number OAC 1920103.

\bibliographystyle{apsrev4-1}
\bibliography{reference}

\clearpage
\appendix
\begin{widetext}
\section{Derivation of the effective model}
\subsection{Wannier state and tight binding model in the bottom layer}
We focus on the case I as an example. Now both the top and bottom layer are doped with holes in the valence band, with $n^h_b=1-x, n^h_t=x$. In the hole picture, the microscopic Hamiltonian describing this moir\'e+monolayer system is
\begin{align}
    H=H_b+H_t+H_{tb},
\label{eq1}
\end{align}

$H_b$ is the Hamiltonian in the bottom layer. The holes in the WSe$_2$ layer feels a moir\'e superlattice:
\begin{align}
H_b^0=&\int d{\bf x}\left(-h_b^\dag({\bf x})\frac{\nabla^2}{2m}h_b({\bf x})+\sum_{{\bf G}_i}V_{{\bf G}_i}({\bf x})h_b^\dag({\bf x})h_b({\bf x})\right)\nonumber\\+&\int d{\bf x}d{\bf x}^\prime \frac{1}{2}V_{bb}({\bf x}-{\bf x}^\prime)\rho({\bf x})\rho({\bf x}^\prime),
\end{align}
here $h_b^\dag({\bf x})$ is the hole creation operator in the bottom layer.  $\rho({\bf x})=h_b^\dag({\bf x})h_b({\bf x})$ is the hole density operator. ${{\bf G}_i}=\frac{4\pi}{\sqrt{3}a_M}(\cos\frac{i2\pi}{3},\sin\frac{i2\pi}{3})$ with $i=0,1,2$. $V_{G}$ is the superlattice potential which is $10meV$ in our calculation and $a_M$ is the moir\'e superlattice constant $\sim\frac{a}{(\delta a)^2+\theta^2}$, where $a$ is the lattice constant of WSe$_2$. $\delta a$ is the mismatch of WSe$_2$ and WS$_2$, and $\theta$ is the twisted angle, which is $0$ in our calculation. $V_{bb}({\bf x}-{\bf x}^\prime)$ is the Coulomb interaction in the bottom layer.\\
\indent The Hamiltonian in the top layer $H_t$ is
\begin{align}
    H_t=-\int d{\bf x}h_{t}^\dag({\bf x})\frac{\nabla^2}{2m}h_t({\bf x})+
    \frac{1}{2}\int d{\bf x}d{\bf x}^\prime V_{tt}({\bf x}-{\bf x}^\prime)n_t({\bf x})n_t({\bf x}^\prime),
\end{align}
where $h_t^\dag({\bf x})$ creates an hole in the top layer at position ${\bf x}$ and $n_t({\bf x})=h_t^\dag({\bf x})h_t({\bf x})$ is the hole density in the top layer. $V_t({\bf x}-{\bf x}^\prime)$ is the Coulomb potential in the top layer. Note that there is no moir\'e superlattice and we have free hole gas.

\indent The inter-layer interaction term $H_{tb}$ consists of the inter-layer Coulomb interaction,
    \begin{align}
    H_{tb}=&\int d{\bf x}d{\bf x}^\prime V_{tb}({\bf x}-{\bf x}^\prime)n^h_b({\bf x})n^h_t({\bf x}^\prime),
\end{align}
where $V_{tb}$ describes the inter-layer Coulomb interaction.\\
\indent The parameters in the our calculation are: lattice constant $0.328nm$ for WSe$_2$, lattice constant $0.315nm$ for WS$_2$, the moir\'e lattice constant is $a_M=7.9nm$, $m$ is $0.42m_e$, corresponding to the kinetic energy $\frac{\hbar^2}{ma_M^2}=2.90703meV$ and potential energy $\frac{1}{4\pi\epsilon_0}\frac{e^2}{a_M}=182.274meV$.
The Coulomb interactions are
\begin{align}
    V_{tt}({\bf q})=&V_{bb}({\bf q})=\frac{e^2}{2\epsilon_0\epsilon|{\bf q}|}(1-e^{-D|{\bf q}|}),\\
    V_{bt}({\bf q})=&\frac{e^2}{2\epsilon_0\epsilon|{\bf q}|}\left(e^{-d|{\bf q}|}-e^{-D|{\bf q}|}\right),
\end{align}
where $d$ and $D$ correspond to distance between the moir\'e layer and the monolayer and the distance between the monolayer and the gate, respectively. We adopt $d=2nm$ and $D=30nm$. In real space they correspond to,
\begin{align}
    V_{tt}({\bf x})=&\frac{e^2}{4\pi\epsilon_0\epsilon}(\frac{1}{|{\bf x}|}-\frac{1}{\sqrt{{\bf x}^2+D^2}}),\\
    V_{bt}({\bf x})=&\frac{e^2}{4\pi\epsilon_0\epsilon}\left(\frac{1}{\sqrt{{\bf x}^2+d^2}}-\frac{1}{\sqrt{{\bf x}^2+D^2}}\right)
\end{align}

We  first try to obtain the moir\'e bands  in the bottom layer using the continuum model:
\begin{align}
    H_b=\sum_{{\bf k}}\frac{\hbar^2{\bf k}^2}{2m_b}h_{\bf k}^\dag h_{\bf k}+\sum_{\mathbf k,\mathbf G}V({\bf k}+\mathbf G_i)(h_{{\bf k}+\mathbf G_i}^\dag h_{\bf k}+h.c),
\end{align}
here $\mathbf G_1=\frac{2\pi}{a_M}(0,\frac{2}{\sqrt{3}})$, $\mathbf G_2=\frac{2\pi}{a_M}(1,-\frac{1}{\sqrt{3}})$,  $\mathbf G_3=\frac{2\pi}{a_M}(-1,\frac{1}{\sqrt{3}})$.

To simplify our analysis, following Ref.~\onlinecite{PhysRevB.99.205150}, we first construct  Wannier orbitals from $H_b$ to derive a low energy effective model. The Wannier state is
\begin{align}
    h_i^\dag=\frac{1}{\sqrt{N}}\sum_k e^{\mathbbm{i}{\bf k}\cdot{\bf R}_i}e^{\mathbbm{i}\theta({\bf k})}h^\dag({\bf k}),
\end{align}
$\theta({\bf k})$ can be calculate from the projection method\cite{RevModPhys.84.1419}, with $e^{\mathbbm{i}\theta({\bf k})}=\frac{\langle\mu({\bf k})|g({\bf k})\rangle}{|\langle\mu({\bf k})|g({\bf k})\rangle|}$. Here $\mu({\bf k})$ is the Bloch wave function with energy $\xi({\bf k})$, and $g({\bf k})$ is trail wave function (in our calculation, it is Gaussian).

In the Wannier orbital basis, the hopping term $t$ is
\begin{align}
    t(m,n)=-\frac{1}{N}\sum_{{\bf k}}\xi({\bf k})e^{-\mathbbm{i}{\bf k}\cdot(m{\bf a}_1+n{\bf a}_2)},
\end{align}

The intra-layer interaction in the bottom layer $\frac{1}{2}V_{bb}(x-x^\prime)\rho_b(x)\rho_b(x^\prime)$ can be rewritten as:
\begin{align}
   \frac{1}{2} \sum_{x,R_1,R_2,R_3}U(R_1,R_2,R_3)h^\dag(x)h^\dag(x+R_1)h(x+R_2)h(x+R_3),
\end{align}
with
\begin{align}
    U(R_1,R_2,R_3)=&\frac{1}{N^2}\sum_{{\bf k}_1,{\bf q}_1,{\bf k}_2,{\bf q}_2}\sum_{x,x^\prime}V_{bt}(x-x^\prime)\lambda({\bf k}_1,{\bf q}_1)\lambda({\bf k}_2,{\bf q}_2)\nonumber\\e&^{-\mathbbm{i}\theta({\bf k}_1+{\bf q}_1)}e^{-\mathbbm{i}\theta({\bf k}_1)}e^{-\mathbbm{i}\theta({\bf k}_2+{\bf q}_2)}e^{-\mathbbm{i}\theta({\bf k}_2)}e^{\mathbbm{i}({\bf q}_1+{\bf q}_2)x^\prime}e^{\mathbbm{i}({\bf k}_2+{\bf q}_2)R_1}e^{\mathbbm{i}k_2R_2}e^{-\mathbbm{i}{\bf k}_1R_3}e^{-\mathbbm{i}{\bf q}_1x}e^{-\mathbbm{i}{\bf q}_2x^\prime},
\end{align}
where $\lambda({\bf k},{\bf q})=\langle\mu({\bf k})|\mu({\bf k}+{\bf q})\rangle$ is the form factor. 

For the inter-layer Coulomb interaction $\int dxdx^\prime V_{bt}(x-x^\prime)n_b(x)n_t(x^\prime)$, we substitute $n_b(x)$ with $\sum_iW_i^*(x-{\bf R}_i)W_i(x-{\bf R}_i)$. It becomes
\begin{align}
    \sum_i\int dxdx^\prime W_i^*(x^\prime-{\bf R}_i)W_i(x^\prime-{\bf R}_i)V_{bt}(x-x^\prime)n_t(x)=\sum_i\int dxV_i(x-{\bf R}_i)n_t(x),
\end{align}
with $V_i(x-{\bf R}_i)=\int dx^\prime W_i^*(x^\prime-{\bf R}_i)W_i(x^\prime-{\bf R}_i)V_{bt}(x-x^\prime)$, where $W_i(x-{\bf R}_i)$ is the Wannier function. The hole in the top layer at $x$ feels a potential $\sum_{i}V_i(x-{\bf R}_i)$. When we use $V_{bt}(x-x^\prime)$, it means we consider the bottom layer as a continuous model with holes located at $x$, while when we use $V_i(x-{\bf R}_i)$, it means we consider the bottom layer as a lattice model with holes located at lattice site ${\bf R}_i$.

\indent The extended lattice Hubbard model describing the bottom layer is:
\begin{align}
    H_b=-t_b\sum_{<ij>}h^\dag_{bi}h_{bj}+\frac{U}{2}\sum_{i}n_{bi}^2+V\sum_{<ij>}n_{bi}n_{bj},
\end{align}
where $h_{bi}^\dag$ creates a hole at site $i$, and $n_{bi}=h_{bi}^\dag h_{bi}$ is the hole density. $t_b=1.06meV$ is the hooping in the bottom layer, and $U=660.634meV/\epsilon$ and $V=92.689meV/\epsilon$ correspond to the on site Hubbard interaction and nearest-neighbor Hubbard interaction in the bottom layer.

\subsection{Exciton wave function and derivation of effective spin-exciton model}
The total Hamiltonian now becomes  
\begin{align}
    H=-&t_b\sum_{<ij>}h_{bi;\sigma}^\dag h_{bj;\sigma}+\frac{U}{2}\sum_{i}n_{bi}^2+V\sum_{<ij>}n_{bi}n_{bj}\nonumber\\
-&\int_x h_{t}^\dag(x)\frac{\nabla^2}{2m}h_t(x)+\frac{1}{2}\int dxdx^\prime V_{tt}(x-x^\prime)n_t(x)n_t(x^\prime)+\sum_i\int dxV_i(x-{\bf R}_i)n_{bi}n_t(x^\prime),
\label{eqb1}
\end{align}
where $n_{bi}=\sum_{\sigma=\uparrow,\downarrow}h^\dagger_{bi;\sigma}h_{bi;\sigma}$.

\subsubsection{One exciton problem}
First we consider there only one hole in the bottom layer, and one electron in the top layer. We can calculate the wave function by solving the Schr\"odinger equation,
\begin{align}
    \left(-\frac{\hbar^2\nabla^2}{2m}+\sum_iV_i(x-R_i)-V_0(x)\right)\phi_{in}(x)=E_n\phi_{in},
\end{align}
where $E_n$ and $\phi_{in}$ correspond to the $n$-th eigenenergy and $n$-th eigenstate, respectively. An excitonic state can be expressed as
\begin{align}
    b^\dag_i|0\rangle=\int_x\phi_{in}(x)h_t^\dag(x)h_{bi}\prod_j h_{bj}^\dag|0\rangle,
\end{align}
here the spin indices are implicit. The exciton hopping term between a nearest neighbor pair $\langle ij \rangle$ is $J_p b_i^\dag b_j$, where $J_p$ can be calculated from the first order perturbation theory:
\begin{align}
    J_p=\langle 0|b_i H_b b_j^\dag|0\rangle=\langle 0|\int_x\phi_{in}^*(x)(\prod_l h_{bl}) h_{bi}^\dag h_{t}(x)(-t_{b}h_{bk}^\dag h_{bl})\int_{x^\prime}\phi_{jn}(x^{\prime})h_t^\dag(x^\prime)h_{bj}\prod_k h_{bk}^\dag|0\rangle=t_{b}\int_x\phi_{in}^*(x)\phi_{jn}(x),
\end{align}
where in the second equality we used the relation $h_{bi}^\dag h_{bk}^\dag h_{bl}h_{bj}=-h_{bi}^\dag  h_{bl}h_{bk}^\dag h_{bj}=-\delta_{il}\delta_{kj}$. Note here we always have $i\neq j$.  Thus the effective hopping of exciton is $J_p=t_{b}\int_x\phi_{in}^*(x)\phi_{jn}(x)$.  For the case I it is always positive.

\subsubsection{Two exciton problem}
Next we calculate $J_t$, which is the energy difference of the triplet state and singlet state $E_t-E_s$ of the two exciton problem. We need to consider two holes in the top layer locating in the nearest neighbor pair $\langle ij \rangle$ and two vacancies of hole (two electrons) in the bottom layer. The Schr\"odinger equation is:
\begin{align}
    \left(-\frac{\hbar^2\nabla^2}{2m}+\sum_iV_i(x-R_i)-V_i(x-R_i)-V_j(x-R_j)\right)\phi_{ij}^n(x)=E_n\phi_{ij}^n,
\end{align}
With $E_n$ the $n$-th eigenenergy, and $\phi_{ij}^n$ the $n$-th eigenstate. With the existence of Coulomb interaction in the top layer, we can construct the symmetric and anti-symmetric wave function using the ground state and first excited state wave function $\phi_{ij}^0$ and $\phi_{ij}^1$,
\begin{align}
    \psi_{+}=&\phi_{ij}^0(x_1)\phi_{ij}^0(x_2),\\
    \psi_{-}=&\frac{1}{\sqrt{2}}(\phi_{ij}^0(x_1)\phi_{ij}^1(x_2)-\phi_{ij}^0(x_2)\phi_{ij}^1(x_1)),\\    
    J_t=&\langle\psi_-|H|\psi_-\rangle-\langle\psi_+|H|\psi_+\rangle,
\end{align}
the spin indices is implicit hereinafter.
\begin{align}
 E_t=&\int \frac{1}{\sqrt{2}}(\phi_{ij}^0(x_1)\phi_{ij}^1(x_2)-\phi_{ij}^0(x_2)\phi_{ij}^1(x_1))^* (H_1+H_2+\frac{1}{2}V_{tt}(x_1-x_2))\frac{1}{\sqrt{2}}(\phi_{ij}^0(x_1)\phi_{ij}^1(x_2)-\phi_{ij}^0(x_2)\phi_{ij}^1(x_1)),\\
 =&E_0+E_1+\frac{1}{2}\frac{1}{2}\int{\phi_{ij}^0(x_1)}^*{\phi_{ij}^1(x_2)}^*\phi_{ij}^0(x_1)\phi_{ij}^1(x_2)V^\prime(x_1-x_2)+{\phi_{ij}^0(x_2)}^*{\phi_{ij}^1(x_1)}^*\phi_{ij}^0(x_2)\phi_{ij}^1(x_1)V_{tt}(x_1-x_2)\\
 -&{\phi_{ij}^0(x_1)}^*{\phi_{ij}^1(x_2)}^*\phi_{ij}^0(x_2)\phi_{ij}^1(x_1)V^\prime(x_1-x_2)-{\phi_{ij}^0(x_2)}^*{\phi_{ij}^1(x_1)}^*\phi_{ij}^0(x_1)\phi_{ij}^1(x_2)V_{tt}(x_1-x_2),\\
 E_s=&2E_0+\frac{1}{2}\int V_{tt}(x_1-x_2){\phi_{ij}^0(x_1)}^*{\phi_{ij}^0(x_2)}^*\phi_{ij}^0(x_1)\phi_{ij}^0(x_2),\\
 J_t=&E_t-E_s.
\end{align}
We calculate $J_t$ from the above equations numerically.

Next we move to estimate the exciton dipole-dipole repulsion $J_{pz}$. The term $J_{pz}(n_t(i)-n_b(i))(n_t(j)-n_b(j))$ is diagonal in the basis $|t,t\rangle,|t,b\rangle,|b,t\rangle,|b,b\rangle$,
\begin{align}J_{pz}
    \begin{pmatrix}
    1&0&0&0\\
    0&-1&0&0\\
    0&0&-1&0\\
    0&0&0&1
    \end{pmatrix}
\end{align}
the basis in our model are expressed as
\begin{align}
    |t,t\rangle=&\int dx_1dx_2\phi_{ij}(x_1)\phi_{ij}(x_2)h^\dag_{t}(x_1)h^\dag_t(x_2)|0\rangle,\\
    |t,b\rangle=&\int dx_1\phi_i(x_1)h^\dag_{t}(x_1)h^\dag_{bj}|0\rangle,\\
    |b,t\rangle=&\int dx_1\phi_j(x_1)h^\dag_{t}(x_1)h^\dag_{bi}|0\rangle,\\
    |b,b\rangle=&h^\dag_{bi}h^\dag_{bj}|0\rangle,
\end{align}
where $\phi_{ij}(x)=\phi_{ij}^0(x)$.

There energies can be calculated as:
\begin{align}
E_{tt}=&\langle t,t|V\sum_{<ij>}n_{bi}n_{bj}+\frac{1}{2}\int dxdx^\prime V_{tt}(x-x^\prime)n_t(x)n_t(x^\prime)+\int dxdx^\prime V_{bt}(x-x^\prime)n_b(x)n_t(x^\prime)|t,t\rangle\nonumber\\
=&\frac{1}{2}
\int dx_1dx_2V_{tt}(x_1-x_2){\phi_{ij}(x_1)}^*{\phi_{ij}(x_2)}^*\phi_{ij}(x_1)\phi_{ij}(x_2)
+\frac{1}{2}\int dx_1dx_2V_{tt}(x_2-x_1){\phi_{ij}(x_1)}^*{\phi_{ij}(x_2)}^*\phi_{ij}(x_1)\phi_{ij}(x_2),
\end{align}
\begin{align}
E_{tb}=&\langle t,b|V\sum_{<ij>}n_{bi}n_{bj}+\frac{1}{2}\int dxdx^\prime V_{tt}(x-x^\prime)n_t(x)n_t(x^\prime)+\int dxdx^\prime V_{bt}(x-x^\prime)n_b(x)n_t(x^\prime)|t,b\rangle\nonumber\\
=&\int dx_1dx_2V_{bt}(x_1-x_2){W_j(x_1-{\bf R}_j)}^*W_j(x_1-{\bf R}_j){\phi_i(x_2)}^*\phi_i(x_2),
\end{align}
\begin{align}
E_{bt}=&\langle b,t|V\sum_{<ij>}n_{bi}n_{bj}+\frac{1}{2}\int dxdx^\prime V_{tt}(x-x^\prime)n_t(x)n_t(x^\prime)+\int dxdx^\prime V_{bt}(x-x^\prime)n_b(x)n_t(x^\prime)|b,t\rangle\nonumber\\
=&\int dx_1dx_2V_{bt}(x_1-x_2){W_j(x_1-{\bf R}_j)}^*W_j(x_1-{\bf R}_j){\phi_j(x_2)}^*\phi_j(x_2),
\end{align}
\begin{align}
E_{bb}=\langle b,b|V\sum_{<ij>}n_{bi}n_{bj}+\frac{1}{2}\int dxdx^\prime V_{tt}(x-x^\prime)n_t(x)n_t(x^\prime)+\int dxdx^\prime V_{bt}(x-x^\prime)n_b(x)n_t(x^\prime)|b,b\rangle=V,
\end{align}

 After getting the diagonal matrix element, together with $n_t=\frac{1}{2}+P_z$ and $n_b=\frac{1}{2}-P_z$, we can calculate $J_{pz}$ by
\begin{align}  E_{tt}n_t(i)n_t(j)+E_{tb}n_t(i)n_b(j)+E_{bt}n_b(i)n_t(j)+E_{bb}n_b(i)n_b(j)=(E_{tt}+E_{bb}-E_{tb}-E_{bt})P_z(i)P_z(j)+\cdots.
\end{align}
We have $\frac{J_{pz}}{2}=E_{tt}+E_{bb}-E_{tb}-E_{bt}$ and we calculate it numerically.

\subsection{The derivation of parameters in case II}
In the last section, we consider the case I where the doped electron and hole reside on the valence bands.  Here we consider the case II with an additional hole to the valence band and an additional electron to the conduction band. So now we have density $n^e_t=x,n^h_b=1+x$.  We can map it to the case I by doing a particle-hole transformation for the moir\'e layer to get $n^e_t=x,n^e_b=1-x$ where $n^e_b$ is the number of electrons in the first moir\'e band on top of the original valence band.  Now the Hamiltonian is exactly the same as the last section except that $t_b$ gets a minus sign from the particle-hole transformation.  All of the calculations follow except now we need to add an additional minus sign to the exciton hopping term $J_p$.  The evolution of parameters with the dielectric constant $\epsilon$ are shown in Fig.~\ref{fig6}(a),
\begin{figure}[htbp]
\centering
\includegraphics[width=1.\textwidth]{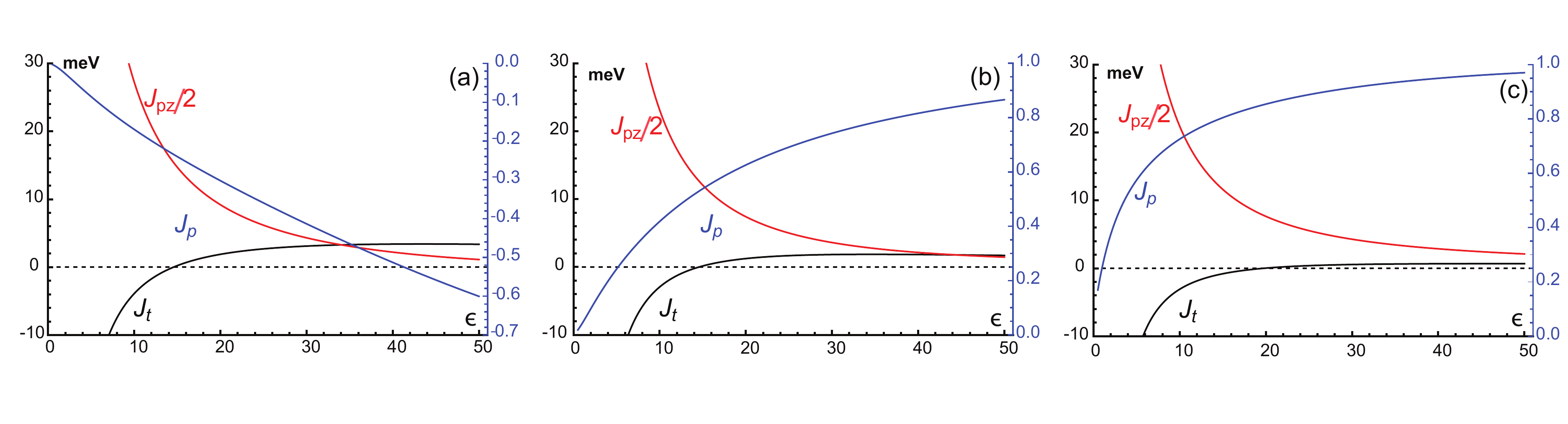}
\caption{The couplings in the effective spin-exciton model as a function of the dielectric constant $\epsilon$. In our calculation, the lattice constants are $0.328nm$ for WSe$_2$, $0.315nm$ for WS$_2$, and the moir\'e lattice constant is $a_0=7.9nm$. $m=0.42m_e$ is the effective mass.  The twist angle $\theta=0^\circ$. The distance is $d=0.1nm$, $d=3nm$, $d=7nm$ for (a), (b), (c) respectively. And In Fig(a), we consider the case II, so $J_p$ is negative. But $J_t,J_{pz}$ are basically the same for the case I and case II.}
\label{fig6}
\end{figure}

\section{reduction to bosonic $t$-$J$ model}
If the spin in the top layer is polarized, the four-flavor spin model will reduce to the bosonic $t$-$J$ model. The boson creation operator is defined as,
\begin{align}
    b^\dag_{i\sigma}=\ket{\sigma}_i\bra{0}_i=h^\dag_{bi\sigma}\int\phi(x-{\bf R}_i)h_{it}(x),
\end{align}
this is just the exciton creation operator, so the bosonic hopping term is the exciton hopping $t=-J_p$. Now the spin operator $\vec{S}_i=\frac{1}{2}h^\dag_{bi\alpha}\vec{\sigma}_{i\alpha\beta}h_{bi\beta}$, so the spin interaction is just the spin interaction $\vec{S}_{b}(i)\cdot\vec{S}_{b}(j)$, $J=J_b$. The bosonic number operator $n_i$ counts the number of holes in the bottom layer. Inserting $P_z(i)=\frac{1}{2}(n_t(i)-n_b(i))$ into the $J_{pz}$ term, we can get $V=\frac{J_{pz}}{8}$. We arrive at the bosonic $t$-$J$ model
\begin{align}
    H=-\sum_{<ij>}t_{ij}b_i^\dag b_j+J\vec{S}(i)\cdot\vec{S}(j)+Vn(i)n(j).
\end{align}
\section{Magnetism in the bottom layer}

\indent To understand the $120^\circ$ order in the bottom layer, we can introduce Schwinger bosons $b_{bi\alpha},b_{ti\alpha}$ and write the model Eq.~\ref{eq2} as in Ref.~\onlinecite{zhang2022}. The $J_p$ term becomes $J_p b_{bi\alpha}^\dag b_{bj\alpha}b_{tj\beta}^\dag b_{ti\beta}$. Become the top layer is ferromagnetic ordered, we can approximate $b_{ti}$ and $b_{ti}^\dag$ as $\langle b_{ti}\rangle=\langle b_{ti}^\dag\rangle=\sqrt{M}$, where $M$ is the related to the magnetic order in the top layer. Now $H$ becomes $J_pM b_{bi\alpha}^\dag b_{bj\alpha}$. The spinon dispersion in the bottom layer is $\epsilon(k)=J_pM(\cos{k_x}+\cos{(-\frac{1}{2}k_x+\frac{\sqrt{3}}{2}k_y)})+\cos{(\frac{1}{2}k_x+\frac{\sqrt{3}}{2}k_y)})+\text{constants}$, where the constant terms comes from the $J_{pz}$ term and the chemical potential. For $J_p>0$, the minimums of the spinon dispersion appear at ${\bf K}$ and ${\bf K}^\prime$. The condensation of spinons at ${\bf Q}={\bf K}$ and ${\bf K}^\prime$ gives rise to the $120^\circ$ order in the bottom layer. While for $J_p<0$, the minimum is located at ${\bf Q}=0$, giving the ferromagnetic order in the bottom layer.

\section{Details of DMRG}
In the DMRG simulation, the $su(4)$ generators are defined as $S_{ab}=\ket{a}\bra{b}$ with $a,b=1,2,3,4$. The operators $\vec{S}$, $\vec{S}_b$, $\vec{S}_t$, $\vec{P}$, can be written as,
\begin{align}
    {S}_{t}^+=&S_{12},{S}_{t}^-=S_{21},S_{t}^z=\frac{1}{2}(S_{11}-S_{22}),\\
    {S}_{b}^+=&S_{34},{S}_{b}^-=S_{43},S_{b}^z=\frac{1}{2}(S_{33}-S_{44}),\\
    {P}^+=&S_{13}+S_{24},{P}^-=S_{31}+S_{42},P_z=\frac{1}{2}(S_{11}+S_{22}-S_{33}-S_{44}),\\
    \vec{S}=&\vec{S}_t+\vec{S}_b
\end{align}
The terms in Eq.~\ref{eq2} is
\begin{align}
    \vec{S}_t(i)\cdot\vec{S}_t(j)=&\frac{1}{2}S_{12}(i)S_{21}(j)+\frac{1}{2}S_{21}(i)S_{12}(j)+\frac{1}{4}(S_{11}(i)S_{11}(j)+S_{22}(i)S_{22}(j)-S_{11}(i)S_{22}(j)-S_{22}(i)S_{11}(j)),\\
    \vec{S}_t(i)\cdot\vec{S}_t(j)=&\frac{1}{2}S_{34}(i)S_{43}(j)+\frac{1}{2}S_{43}(i)S_{34}(j)+\frac{1}{4}(S_{33}(i)S_{33}(j)+S_{44}(i)S_{44}(j)-S_{33}(i)S_{44}(j)-S_{44}(i)S_{33}(j)),\\
    P_z(i)P_z(j)=&\frac{1}{4}(S_{11}(i)+S_{22}(i)-S_{33}(i)-S_{44}(i))(S_{11}(j)+S_{22}(j)-S_{33}(j)-S_{44}(j)),
\end{align}
\begin{align}
    P^+(i)S^+(i)=&(S_{13}+S_{24})(S_{12}+S_{34})=S_{14},\\
    P^+(i)S^-(i)=&(S_{13}+S_{24})(S_{21}+S_{43})=S_{23},\\
    P^+(i)S^z(i)=&(S_{13}+S_{24})\frac{1}{3}(S_{11}-S_{22}+S_{33}-S_{44})=\frac{1}{2}(S_{13}-S_{24}),\\
    P^-(i)S^+(i)=&(S_{31}+S_{42})(S_{12}+S_{34})=S_{32},\\
    P^-(i)S^-(i)=&(S_{31}+S_{42})(S_{21}+S_{43})=S_{41},\\
    P^-(i)S^z(i)=&(S_{31}+S_{42})\frac{1}{3}(S_{11}-S_{22}+S_{33}-S_{44})=\frac{1}{2}(S_{31}-S_{42}),\\
\end{align}
\begin{align}
    (&P_x(i)P)x(j)+P_y(i)P_y(i))(4\vec{S}(i)\cdot\vec{S}(j)+S_0(i)S_0(j))\nonumber\\
    =&S_{14}(i)S_{41}(j)+S_{23}(i)S_{32}(j)+S_{13}(i)S_{31}(j)+S_{24}(i)S_{42}(j)+(i\leftrightarrow j).
\end{align}
For simplicity, like Ref.~\onlinecite{zhang2022}, we simulate the new Hamiltonian $\tilde{H}=H+\frac{1}{8}J_{pz}\sum_{<ij>}n(i)n(j)$.\\

\section{More DMRG results}
\subsection{DMRG for $J_t=J_b=0$, $J_{pz}=5$}
In Fig.~\ref{fig7}, we plot the evolution of $\langle \vec{S}_b({\bf q})\vec{S}_b(-{\bf q})\rangle$ (Fig.~\ref{fig7}(a)(b)(c)) and $\langle \vec{S}_t({\bf q})\vec{S}_t(-{\bf q})$ (Fig.~\ref{fig7}(d)(e)(f)) as we increase the exciton density $x$ for $J_t=J_b=0$, $J_p=1$ and $J_{pz}=5$. We can see, both the AFM in the bottom layer and the FM in the top layer will eventually disappear after we increase the exciton density. Similar plots in Fig.~\ref{fig9} for $J_t=J_b=0$, $J_p=-1$ and $J_{pz}=5$ show that the FM in both layers are quite robust to exciton density $x$ for the case II.
\begin{figure}[H]
\centering
\includegraphics[width=1\textwidth]{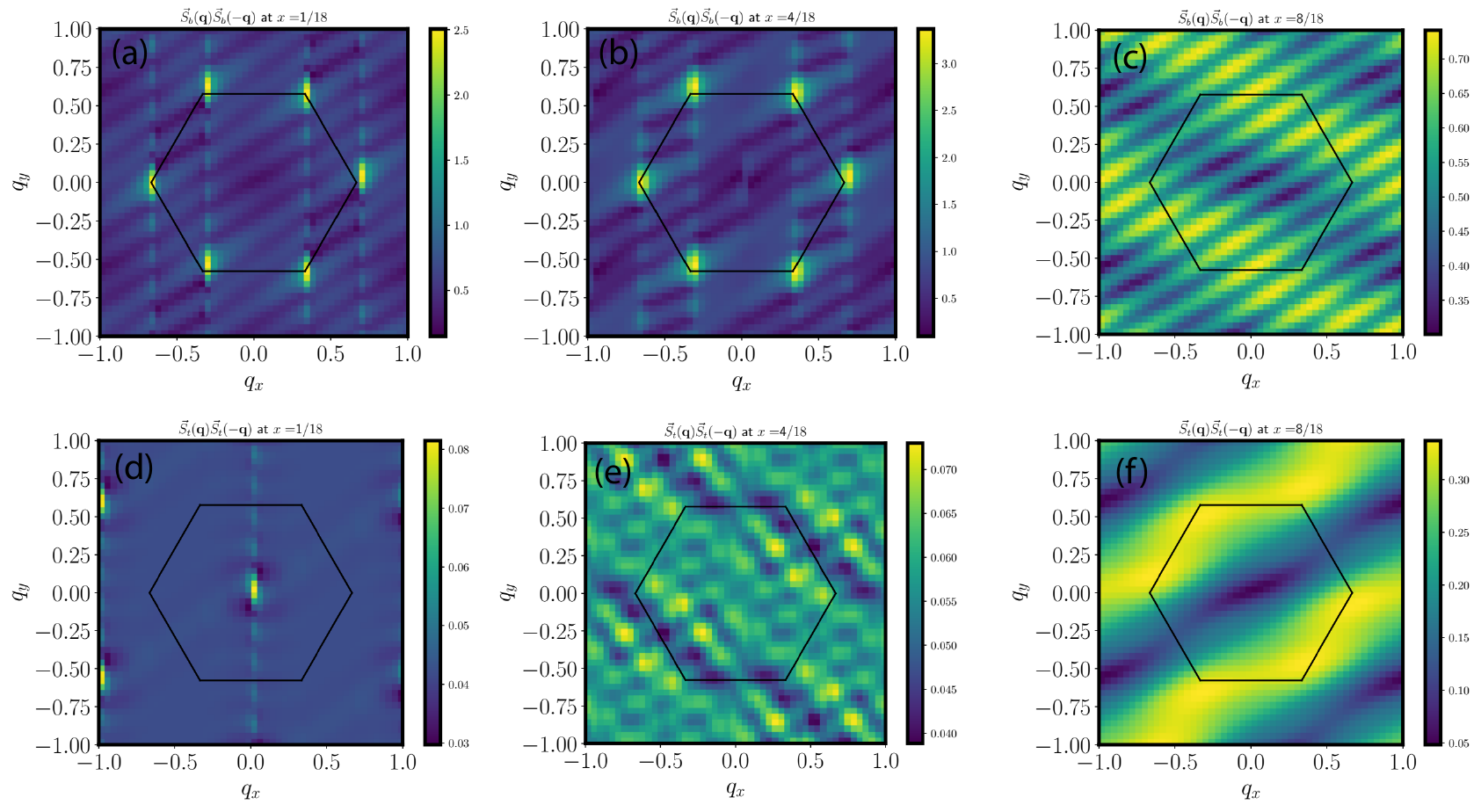}
\caption{$\langle\vec{S}_b({\bf q})\vec{S}_b(-{\bf q})\rangle$ and $\langle\vec{S}_t({\bf q})\vec{S}_t(-{\bf q})\rangle$ for $J_t=J_b=0$, $J_p=1$ and $J_{pz}=5$ at different exciton densities.}
\label{fig7}
\end{figure}
\begin{figure}[H]
\centering
\includegraphics[width=1\textwidth]{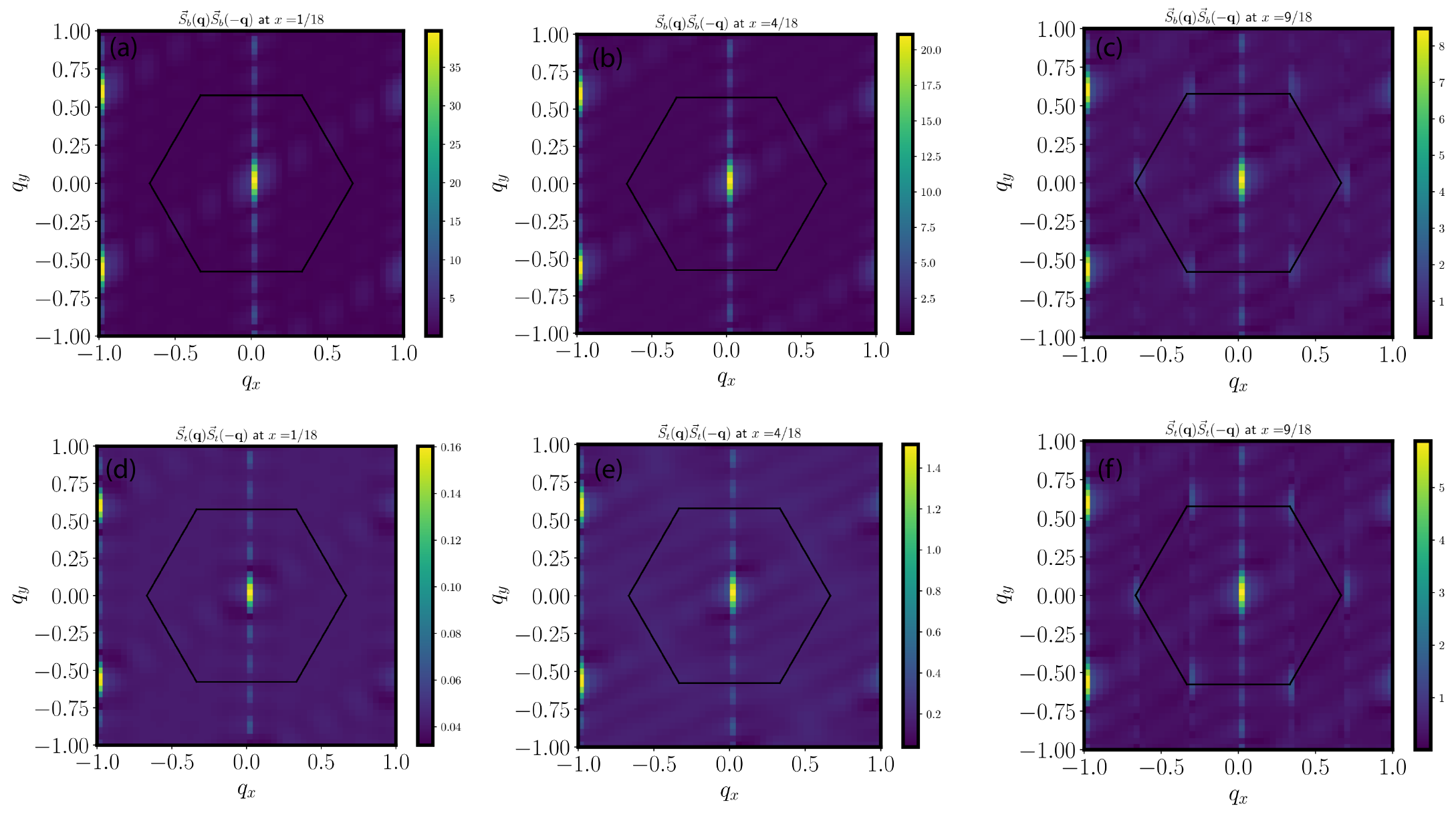}
\caption{$\langle\vec{S}_b({\bf q})\vec{S}_b(-{\bf q})\rangle$ and $\langle\vec{S}_t({\bf q})\vec{S}_t(-{\bf q})\rangle$ for $J_t=J_b=0$, $J_p=-1$ and $J_{pz}=5$.}
\label{fig9}
\end{figure}
In Fig.~\ref{fig8}, we show the exciton correlation for $J_t=J_b=0,J_{pz}=5$ and $J_p=1$ (Fig.~\ref{fig8}(a)(b)(c)), $J_p=-1$ (Fig.~\ref{fig8}(d)(e)(f)).  In the case I with $J_p=1$, when $\vec S_b$ is in the $120^\circ$ AFM phase, the exciton also condenses at the momentum $K,K'$. The exciton condensation is destroyed at larger $x$ when the AFM in $\vec S_b$ is destroyed. In contrast, in the case II with $J_p=-1$, the FM in both layers are robust to exciton density $x$ and the exciton always condenses at $Q=0$. This is a robust spin polarized superfluid phase of excitons.
\begin{figure}[H]
\centering
\includegraphics[width=1\textwidth]{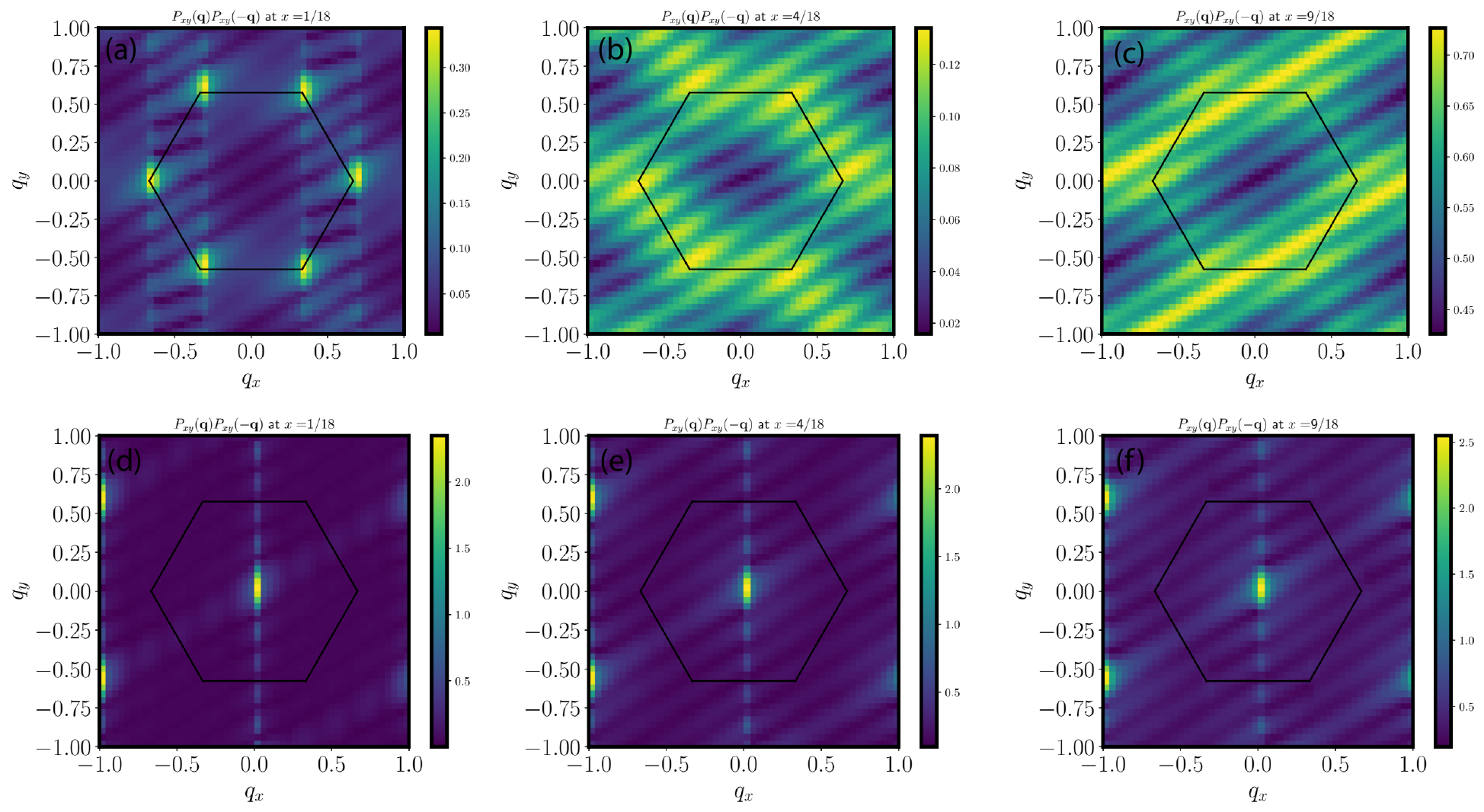}
\caption{$\langle P^\dag({\bf q})P^{-}(-{\bf q})\rangle$ $J_t=J_b=0$, and $J_{pz}=5$. $J_p=1$ (a)(b)(c), and $J_p=-1$ (d)(e)(f).}
\label{fig8}
\end{figure}

\subsection{DMRG for the AFM to FM evolution of case II}

In the case II with $n^e_t=x,n^h_b=1+x$, there is an AFM to FM transition with increasing the exciton density $x$ if there is a finite but small $J_b$. Fig.~\ref{fig11} shows the exciton correlation function of antiferromagnetic to ferromagnetism evolution. When $\vec S_b$ in the AFM phase at small $x$, the exciton does not condense in a commensurate momentum. But when we enter the FM phase, the exciton condenses at the momentum $\mathbf Q=0$. This is a superfluid phase of excitons with both $\vec S_t$ and $\vec S_b$ polarized.

\begin{figure}[H]
\centering
\includegraphics[width=1\textwidth]{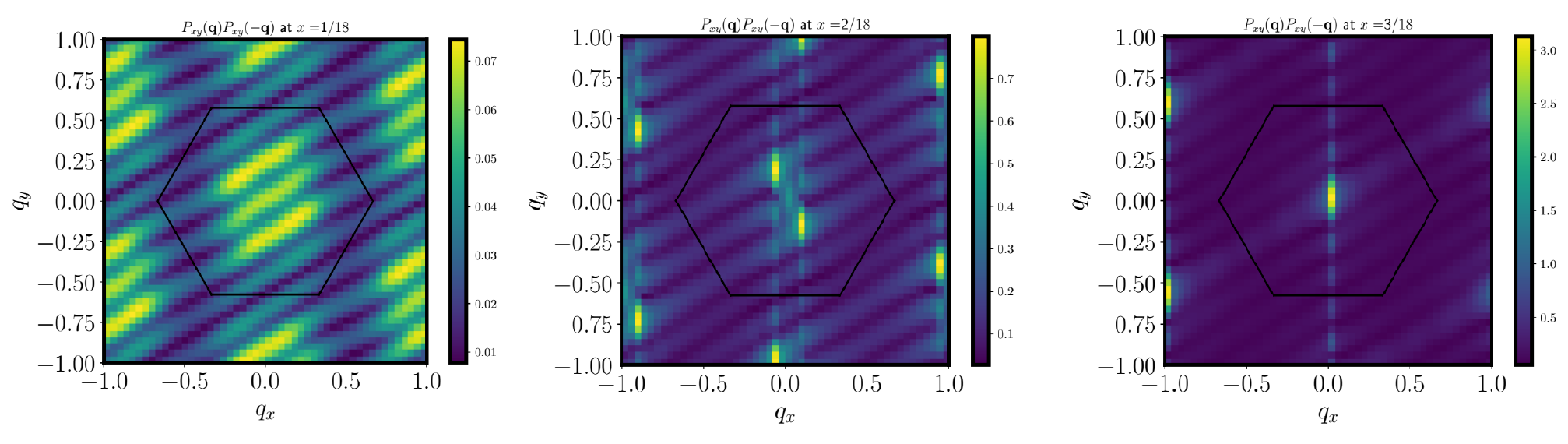}
\caption{$\langle P^\dag({\bf q})P^{-}(-{\bf q})\rangle$ $x=\frac{1}{18}$(a),  $x=\frac{2}{18}$(b) and $x=\frac{1}{18}$(c), the parameters are the same as that in Fig.~\ref{fig5} in the main text for fixed $J_t=0$, $J_{pz}=5$, $J_p=-1$, $J_b=0.06$.}
\label{fig11}
\end{figure}
In Fig.~\ref{fig10}, we show the  spin correlation function  in the bottom layer at $x=\frac{1}{18},\frac{2}{18}$ for fixed $J_t=0$, $J_{pz}=5$, $J_p=-1$. At $x=\frac{2}{18}$, it seems that $\vec S_b$ orders at a small momentum, leading to a spiral phase. At slightly larger $x$ it will become fully spin polarized. At $x=\frac{1}{18}$, the $120^\circ$ AFM is just destroyed, but it does not seem to be in an ordered phase.  This suggests that the AFM to FM evolution goes through an intermediate regime, which we leave to future work for a detailed understanding.

\begin{figure}[ht]
\centering
\includegraphics[width=0.7\textwidth]{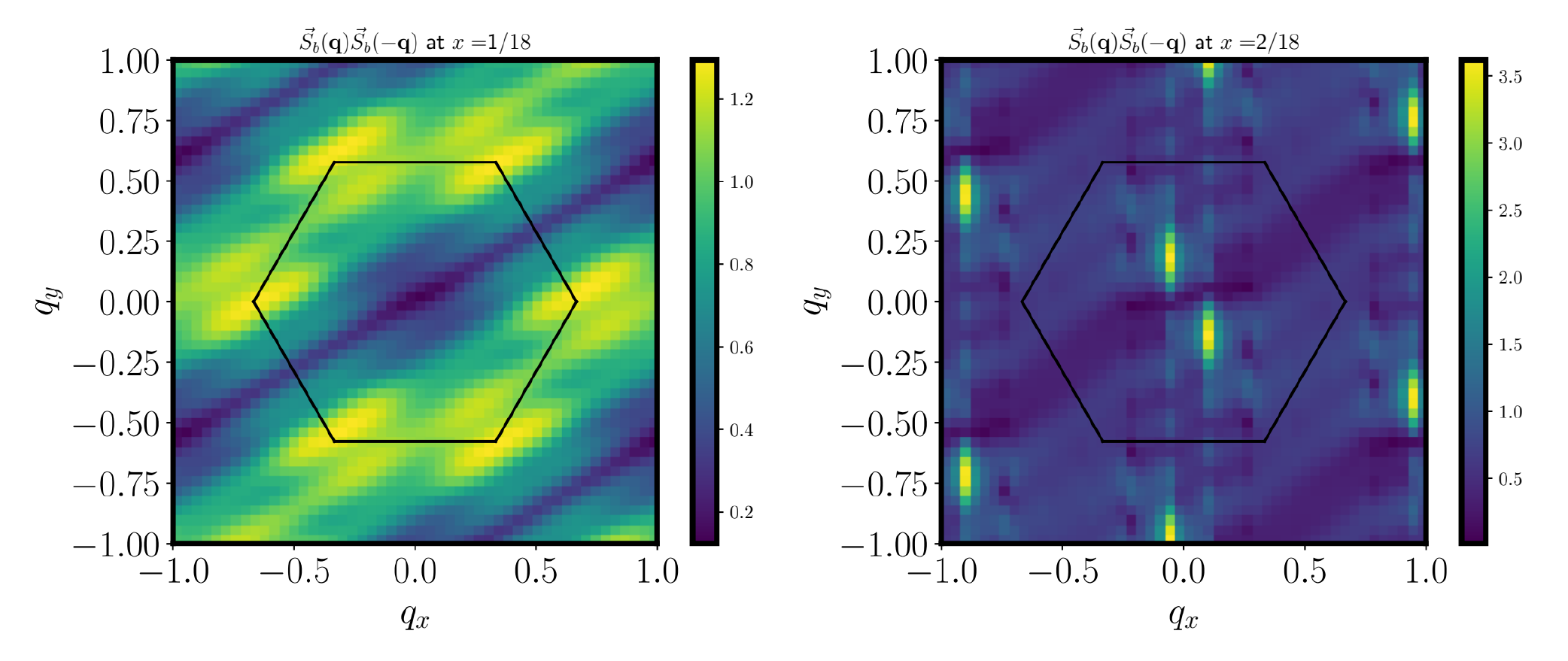}
\caption{Results for $x=\frac{1}{18}$ and $x=\frac{2}{18}$, the parameters are the same as that in Fig.~\ref{fig5} in the main text for fixed $J_t=0$, $J_{pz}=5$, $J_p=-1$, $J_b=0.06$.}
\label{fig10}
\end{figure}

In Fig.~\ref{StSt_Jb=0.06}, we show the  spin correlation function in the top layer for $J_p=-1$, $J_{pz}=5$, and $J_b=0.06$. We find that $\vec S_t$ is always polarized when $\vec S_b$ goes through the AFM to FM evolution.

\begin{figure}[ht]
\centering
\includegraphics[width=0.7\textwidth]{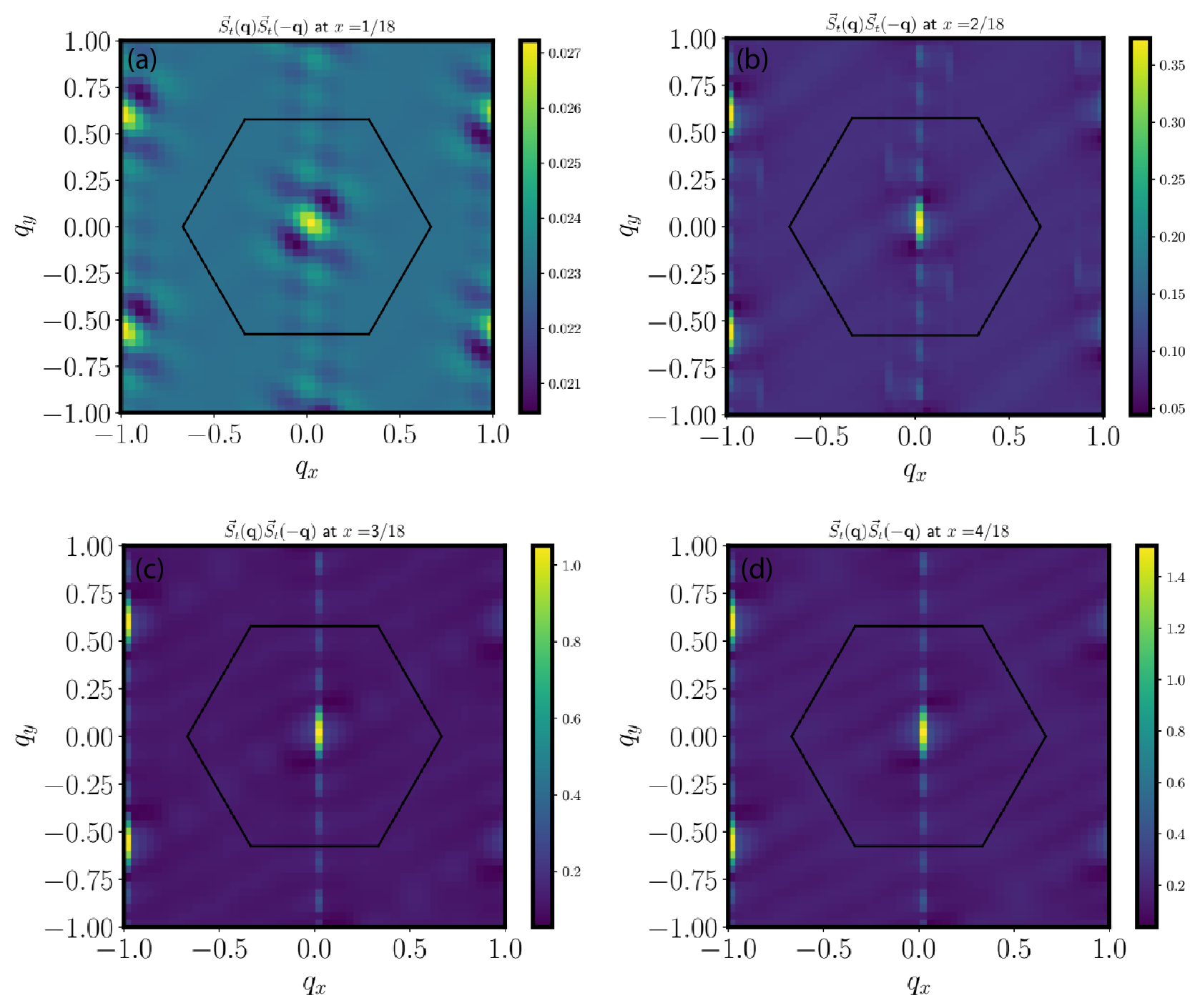}
\caption{Results for $x=\frac{1}{18}$ and $x=\frac{2}{18}$, the parameters are the same as that in Fig.~\ref{fig5} in the main text for fixed $J_t=0$, $J_{pz}=5$, $J_p=-1$, $J_b=0.06$.}
\label{StSt_Jb=0.06}
\end{figure}

\subsection{Robustness of the magnetic order  at large $J_{pz}$}
As shown in our calculation, $J_{pz}$ can be very large. Here, in Fig.~\ref{fig14} and Fig.~\ref{fig15}, we show the result for large $J_{pz}=30$, and we find that the phases we get in the main text is stable for large $J_{pz}$.

\begin{figure*}[htbp]
\centering
\includegraphics[width=0.8\textwidth]{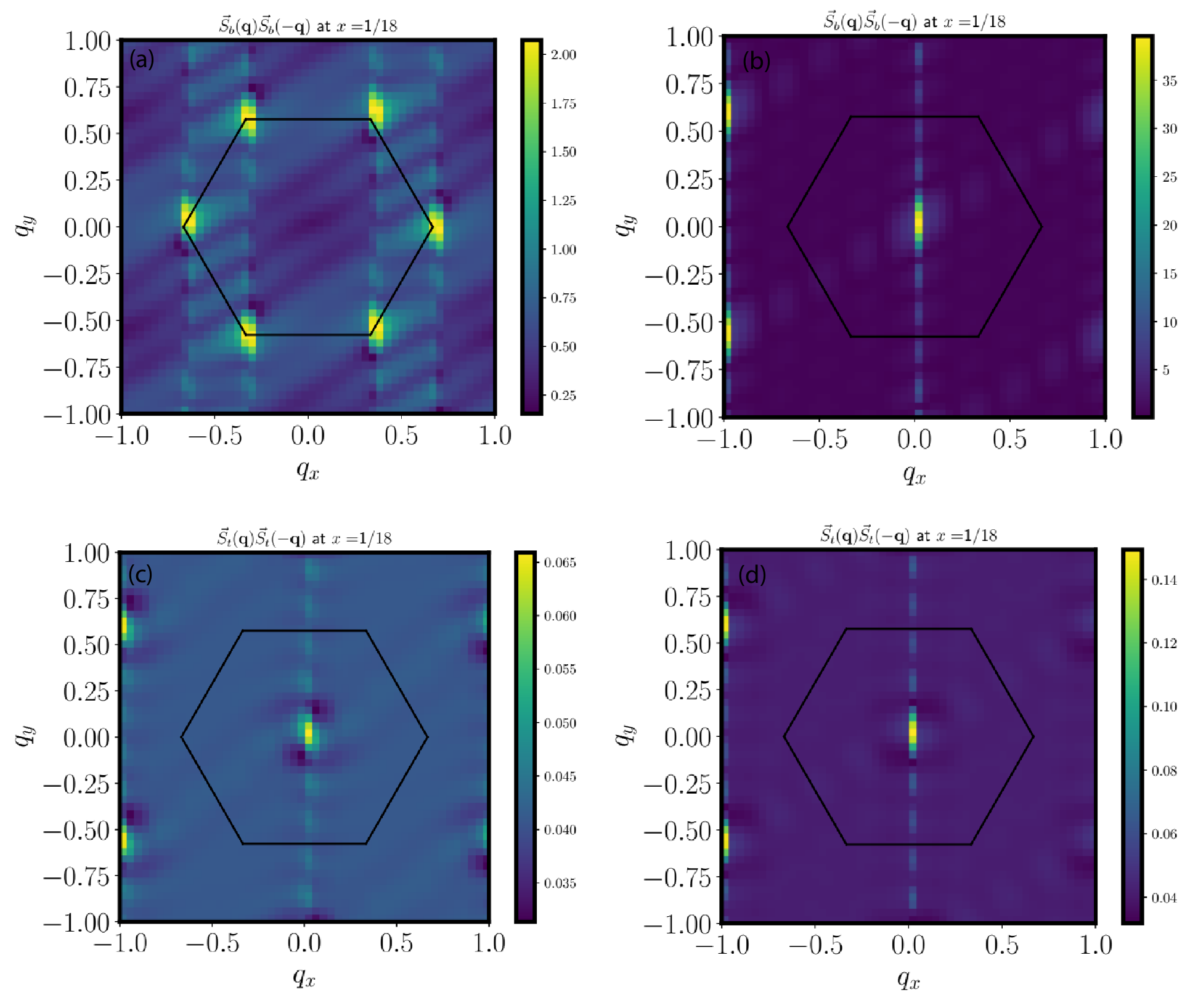}
\caption{The spin correlation function for $J_{pz}=30$, $J_t=J_b=0$, $x=\frac{1}{18}$. (a) $\langle \vec{S}_b({\bf q})\vec{S}_b({-\bf q})\rangle$ for $J_p=1$, (b) $\langle \vec{S}_b({\bf q})\vec{S}_b({-\bf q})\rangle$ for $J_p=-1$, (c) $\langle \vec{S}_t({\bf q})\vec{S}_t({-\bf q})\rangle$ for $J_p=1$, (d) $\langle \vec{S}_t({\bf q})\vec{S}_t({-\bf q})\rangle$ for $J_p=-1$}
\label{fig14}
\end{figure*}

\begin{figure}[H]
\centering
\includegraphics[width=1\textwidth]{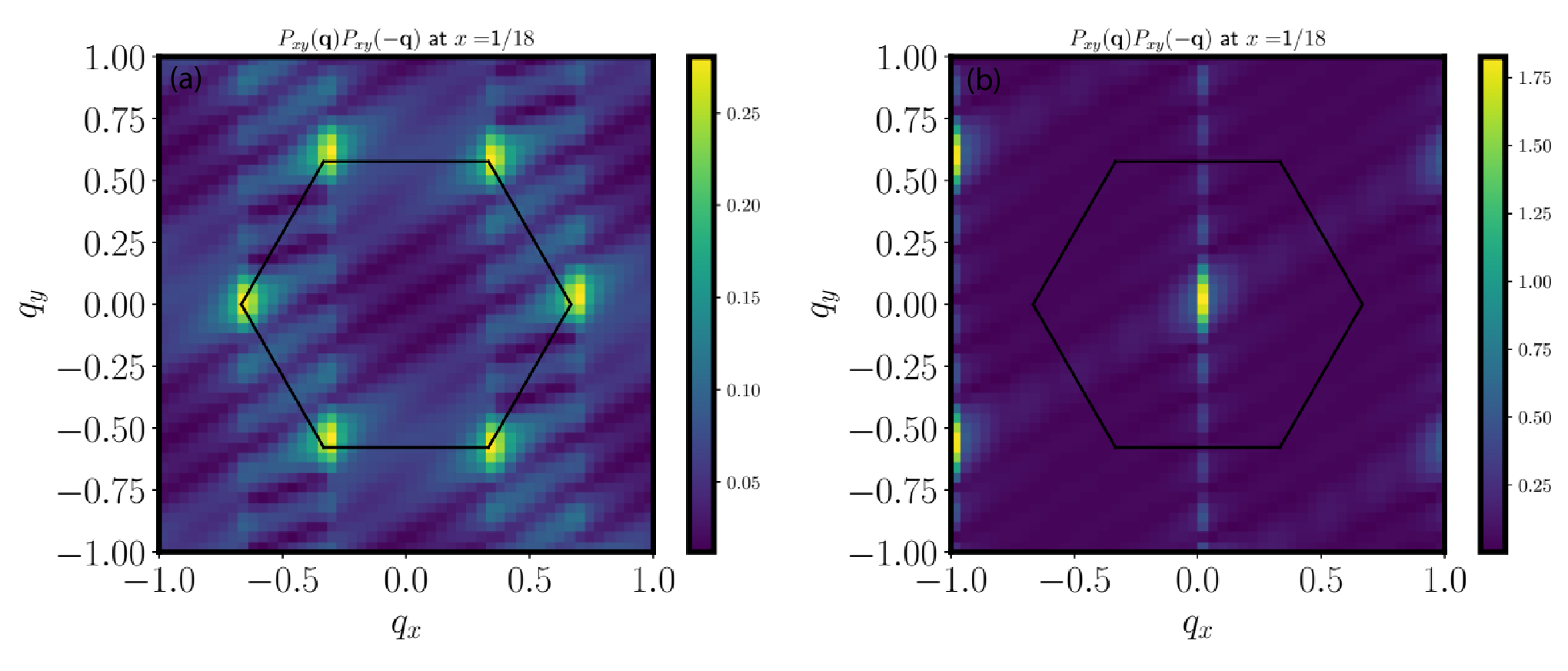}
\caption{The exciton correlation function for $J_{pz}=30$, $J_t=J_b=0$, $x=\frac{1}{18}$. (a), (b) $\langle P^\dag({\bf q}) P^-({\bf q})\rangle$ for $J_p=1$ and $J_p=-1$, respectively.}
\label{fig15}
\end{figure}
\section{Light induced FM at $n=-\frac{1}{3}$}

The experiment\cite{wang2018colloquium} observed light induced ferromagnetism at $n=-\frac{1}{3}$. When the nearest-neighbor interaction is strong enough, the system at $n=-\frac{1}{3}$ is a Wigner crystal with one hole per-triangle in the initial triangular lattice. We can regard it as a new triangular lattice formed by the next-nearest neighbor sites. Now we should derive the new coupling parameters in our model Eq.~\ref{eq2} in the main text. Starting with the tight-binding model in the bottom layer
\begin{align}
    H_b=-t_b\sum_{<ij>}h^\dag_{bi\sigma}h_{bj\sigma}+\frac{U}{2}\sum_{i}n_i^2+V\sum_{<ij>}n_in_j+V_2\sum_{<<ij>>}n_in_j,
\end{align}
where $U=660.635meV/\epsilon$, $V=92.689meV/\epsilon$, $V_2=16.714meV/\epsilon$ correspond to the onsite, nearest, next-nearest Hubbard interaction. We have the exciton state defined as
\begin{align}
    b^\dag_i\ket{0}=\int dx\phi_i(x)c_t^\dag(x)h_{bi}^\dag\ket{0},
\end{align}
now the exciton hopping correspond to second order process. First the exciton (hole) hops to an intermediate site, the first order perturbed wave function is (step $1$ to step $2$ in Fig.~\ref{n_0.3}(b))
\begin{align}  \ket{\psi}=\ket{\psi_0}+\sum_i\ket{\psi_i}\frac{\langle\psi_i|H_1|\psi_0\rangle}{E_0-E_i}=\ket{\psi_0}+\ket{\psi^\prime},
\end{align}
with $\ket{\psi_0}=b^\dag_i\ket{0}$, and then we considet the exciton hops on the nearest-neighbor sites.
\begin{align}
    J_p=&\langle 0|b_i(-t_b h_{bm}^\dag h_{bl})|\psi^\prime\rangle=\int dx^\prime\phi^*_i(x^\prime)h_{bi}c_t(x^\prime)(-t_b h_{bm}^\dag h_{bl})\frac{1}{V_2-V}(-t_b h_{bl}^\dag h_{bk})\int dx\phi_j(x)c_t^\dag(x)h_{bj}^\dag\ket{0}\nonumber\\=&-2\frac{t_b^2}{V-V_2}\int dx\phi^*_i(x)\phi_j(x),
\end{align}
where the factor of $2$ comes from the summation over the site $l$ on the initial triangular lattice.
$J_t$ and $J_{pz}$ can be calculated from the same process as the case for $n=1$. To calculate $J_b$, we need to consider the fourth order process (step 1 to step 5 in Fig.~\ref{n_0.3}(b)),
\begin{align}
   \frac{-t_b}{V_2-V}h_{bi\gamma}^\dag h_{bk\gamma} \frac{t_b^2}{V-U}(h^\dag_{bk\sigma^\prime}h_{bj\sigma^\prime}h^\dag_{bj\sigma}h_{bk\sigma}+h^\dag_{bj\sigma^\prime}h_{bk\sigma^\prime}h^\dag_{bk\sigma}h_{bj\sigma})\frac{-t_b}{V_2-V}h_{bk\beta}^\dag h_{bi\beta}+({i\leftrightarrow j}),
\end{align}
which gives the term $J_b\vec{S}_b(i)\cdot\vec{S}_b(j)$, with $J_b=\frac{16t_b^4}{(V_2-V)^2(U-V)}$, where a factor of $2$ comes from summation over all sites $k$ on the initial triangular lattice, while the other factor of $2$ comes from the exchange of $i$ and $j$. In Fig.~\ref{n_0.3}(c), we shown the evolution of parameter $J_{pz}$, $J_t$ and $J_p$. Then we can apply our result of AFM to FM transition at $n=1$ to $n=\frac{1}{3}$.

\begin{figure}[H]
\centering
\includegraphics[width=1\textwidth]{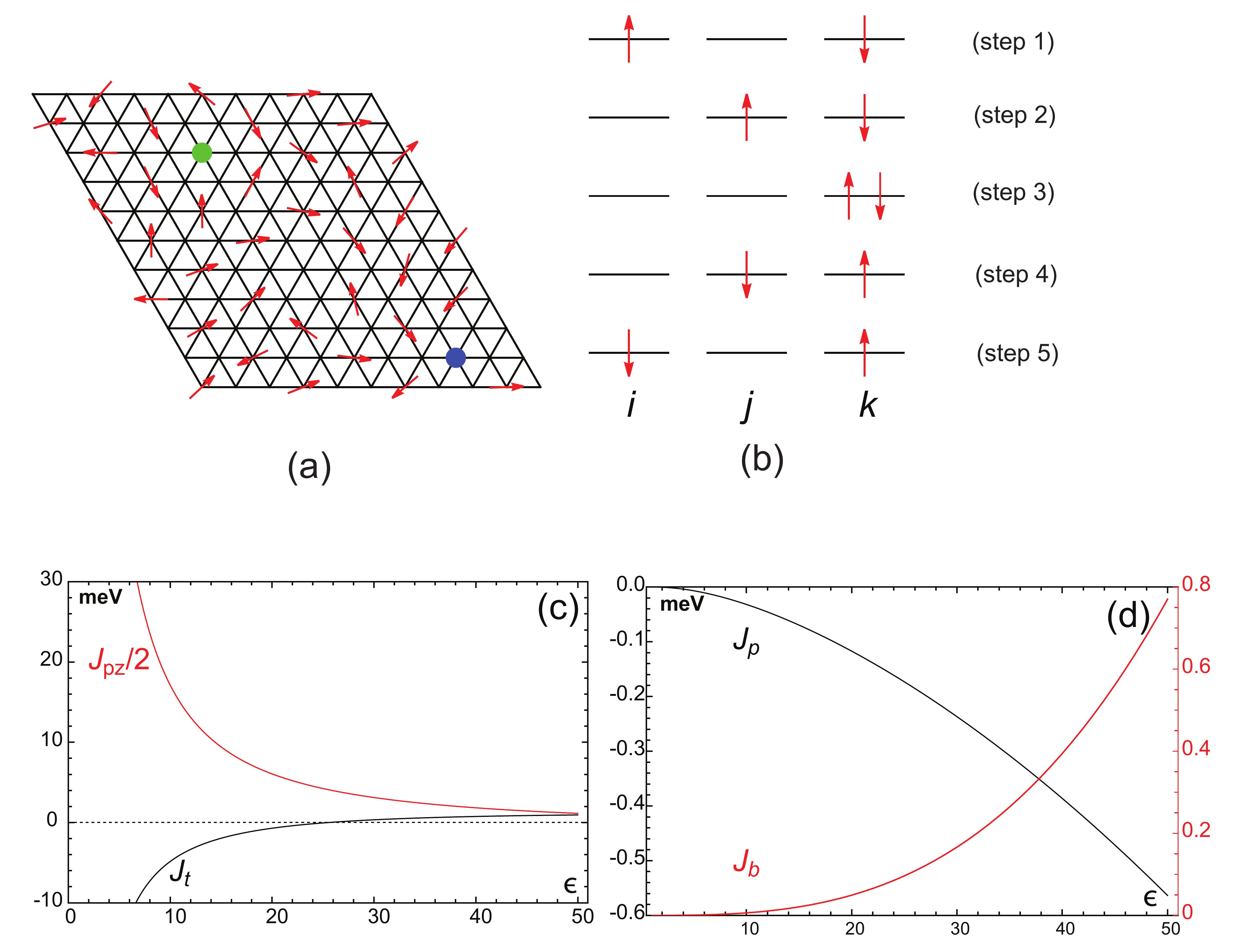}
\caption{(a) Exciton doped Winger crystal at $n=\frac{1}{3}$. (b) perturbation process to generate the spin interaction in the bottom layer. (c) and (d) evolution of parameters. In Fig.(d) the vertical axis on the left and right correspond to $J_p$ and $J_b$, respectively.}
\label{n_0.3}
\end{figure}
\end{widetext}

\end{document}